\documentclass[letterpaper]{article} % DO NOT CHANGE THIS
\usepackage{aaai24}  % DO NOT CHANGE THIS
\usepackage{times}  % DO NOT CHANGE THIS
\usepackage{helvet}  % DO NOT CHANGE THIS
\usepackage{courier}  % DO NOT CHANGE THIS
\usepackage[hyphens]{url}  % DO NOT CHANGE THIS
\usepackage{graphicx} % DO NOT CHANGE THIS
\usepackage{natbib}  % DO NOT CHANGE THIS AND DO NOT ADD ANY OPTIONS TO IT
\usepackage{caption} % DO NOT CHANGE THIS AND DO NOT ADD ANY OPTIONS TO IT
\usepackage{algorithm}
\usepackage{algorithmic}

\usepackage{newfloat}
\usepackage{listings}
\DeclareCaptionStyle{ruled}{labelfont=normalfont,labelsep=colon,strut=off} % DO NOT CHANGE THIS
\floatstyle{ruled}
\newfloat{listing}{tb}{lst}{}
\floatname{listing}{Listing}
\title{Deep Hierarchical Video Compression}
\author{
    Ming Lu\textsuperscript{\rm 1},
    Zhihao Duan\textsuperscript{\rm 2},
    Fengqing Zhu\textsuperscript{\rm 2},
    and Zhan Ma\textsuperscript{\rm 1}\thanks{Corresponding Author}
}
\affiliations{
    \textsuperscript{\rm 1}Nanjing University, Nanjing, Jiangsu, China\\
    \textsuperscript{\rm 2}Purdue University, West Lafayette, Indiana, U.S.\\
    minglu@nju.edu.cn, duan90@purdue.edu, zhu0@purdue.edu, mazhan@nju.edu.cn
}

\usepackage{bibentry}
\usepackage{subfig}
\usepackage{booktabs}
\usepackage{amsmath}
\usepackage{color}
\usepackage{xcolor}
\usepackage{amssymb}
\usepackage{afterpage}

\begin{document}

\maketitle

\begin{abstract}
Recently, probabilistic predictive coding that directly models the conditional distribution of latent features across successive frames for temporal redundancy removal has yielded promising results.  Existing methods using a single-scale Variational AutoEncoder (VAE) must devise complex networks for conditional probability estimation in latent space, neglecting multiscale characteristics of video frames. Instead, this work proposes hierarchical probabilistic predictive coding, for which hierarchal VAEs are carefully designed to characterize multiscale latent features as a family of flexible priors and posteriors to predict the probabilities of future frames. Under such a hierarchical structure, lightweight networks are sufficient for prediction. The proposed method outperforms representative learned video compression models on common testing videos and demonstrates computational friendliness with much less memory footprint and faster encoding/decoding. Extensive experiments on adaptation to temporal patterns also indicate the better generalization of our hierarchical predictive mechanism. Furthermore, our solution is the first to enable progressive decoding that is favored in networked video applications with packet loss.
\end{abstract}

\section{Introduction}
Deep learning breathes fresh life into the visual signal (e.g., images and videos) compression community that has been dominated by handcrafted codecs for decades~\cite{wallace1991jpeg, marcellin2000overview, wiegand2003overview, sullivan2012overview, bross2021overview}.
% Instead of manually designing and optimizing individual modules (e.g., transform, entropy coding, etc.) in traditional compression methods, recent data-driven approaches typically apply end-to-end learning to maximize the overall rate-distortion measures, in which conventional modules are replaced using stacked convolutional or self-attention layers~\cite{lu2022high,li2022hybrid}.
Instead of manually designing and optimizing individual modules such as transforms, mode selection, and quantization in traditional codecs, data-driven approaches adopt end-to-end learning of neural networks~\cite{balle2016end, theis2017lossy}.
%These approaches can be viewed from two different perspectives: non-linear transform coding~\cite{balle2021nonlinear} or variational autoencoders (VAEs)~\cite{balle2018variational}, both of which are much simpler in design than traditional codecs.
Despite the conceptual simplicity, learned image compression methods have achieved superior rate-distortion performance, surpassing the latest VVC (Versatial Video Coding~\cite{bross2021overview}) intra codec~\cite{he2022elic,9810760}.

% even presenting superior performance to the latest, standard compliant codecs~\cite{lu2022high,li2022hybrid}.
% The prominent learned image compression methods~\cite{balle2016end, minnen2018joint} mainly utilized the popular probabilistic model, i.e., Variational Auto-Encoder (VAE) to precisely characterize the data distribution for compression. 

\begin{figure}[t]
\centering
\subfloat[]{\includegraphics[width=0.42\linewidth]{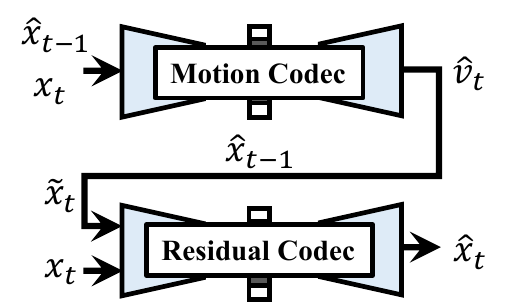}\label{sfig:motion_residual_coding}}
\hspace{0.5cm}
\subfloat[]{\includegraphics[width=0.42\linewidth]{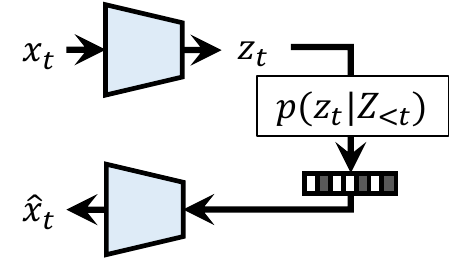}\label{sfig:single-scale_prob_predictive_coding}} \\
\subfloat[]{\includegraphics[width=0.7\linewidth]{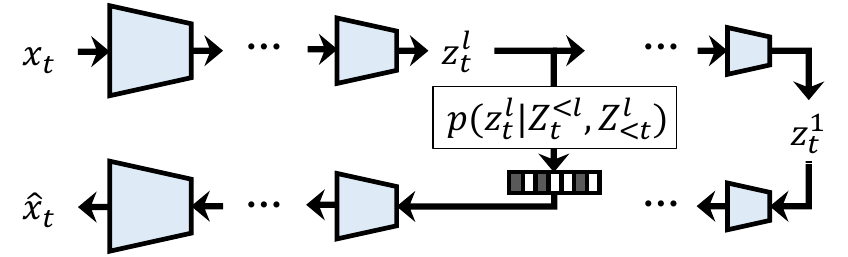}\label{sfig:hier_prob_pred_coding}}
\caption{{Interframe compression in video using (a) {\bf hybrid motion \& residual coding}, (b) {\bf single-scale probabilistic predictive coding}, and (c) {\bf hierarchical probabilistic predictive coding} (Ours).}}
\vspace{-0.2cm}
\end{figure}

For videos, however, learning-based methods are still not free from the shackles of the traditional hybrid framework.
Most existing methods follow the two-stage pipeline shown in Fig.~\ref{sfig:motion_residual_coding}: code motion flows first and then the residual between the current and motion-warped frame, either in an explicit~\cite{lu2019dvc} or conditional~\cite{li2021deep} manner.
This framework is usually cumbersome in design (for example, separate models for intraframe coding, inter residual coding, motion coding, and motion estimation are required); thus, extensive hyperparameter tuning is necessary. 
Furthermore, inaccurate motion-induced warping error propagates inevitably across temporal frames, gradually degrading the quality of reconstructed frames over time.
%it inevitably introduces error propagation across frames, and the two-stage pipeline and post-processing modules cause a high decoding delay (since one has to completely decode the previous frame before starting to decode the current frame).
% the motion compensation module will often result in blurry reconstructions. 
% Emerging learned video compression~\cite{lu2019dvc} are directly evolved from the LICs with rapid development, even outperform the latest VVC recently~\cite{li2022hybrid}. However, those methods are still not free from the shackles of the traditional hybrid framework.
% Most of them follow the two-stage coding pipeline: code the estimated motion flows firstly and then compress the residual between the last frame and the aligned one using the decoded flows in an explicit~\cite{lu2019dvc} or conditional~\cite{li2021deep} manner.
% This framework relies heavily on the accuracy of the estimated motion information with compression noise. It would inevitably introduce the error propagation across the sequence frames and the motion compensation module will often result in blurry reconstructions. In addition, the two-stage coding pipeline and post-processing filtering module will cause unavoidable decoding delay, which has a great negative effect on practical application.

As a promising solution to the problems mentioned earlier, (latent-space)  probabilistic predictive coding attempts to reduce temporal redundancy by conditionally predicting future frames in a one-shot manner.
%without explicit motion estimation and compensation.
% Predictive coding is an important branch of video compression, which attempts to reduce temporal redundancy by only predicting future frames to achieve inter-frame compression.
Intuitively, if the current frame can be well predicted through the past frames,  motion (e.g., flow) estimation and compensation can be completely exempted, and the aforementioned error propagation can also be eliminated accordingly. Recently, Mentzer {\it et al.}~\cite{mentzer2022vct} proposed a probabilistic predictive video coding framework named \textit{Video Compression Transformer (VCT)}. Under the VAE-based image compression framework, VCT models the latent features of the current frame conditioned on the previous-frame latent features using a transformer-based temporal entropy model.
% Though an elegantly simple transformer-based approach to neural video compression, outperforming previous methods without relying on architectural priors such as explicit motion prediction or warping, they rely heavily on the representation ability of the single-scale latent features (i.e., 1/16 resolution of the original frame), constrained the modeling of temporal information.
% In addition, they implemented the Transformer-based prediction according to the autoregressive token-by-token in block size. This approach did not make good use of global structure information for a better feature prior.   
Though VCT outperforms many previous video coding methods, its conditional prediction of single-scale latent features at $1/16$ resolution of the original frame in Fig.~\ref{sfig:single-scale_prob_predictive_coding} fundamentally constrains its characterization capacity, which ignores multiscale characteristics of video frames.

This paper proposes a hierarchical probabilistic predictive coding, termed DHVC, in which conditional probabilities of multiscale latent features of future frames are effectively modeled using deliberately-designed, powerful hierarchical VAEs.
%by carefully using hierarcal VAEs
%In this paper, we build on a more powerful generative model, i.e., ResNet VAE~\cite{kingma2016iafvae, child2021vdvae}, by introducing hierarchical probabilistic modeling of latent variables. 
The latent distribution at a certain scale in the current frame is predicted by the prior features from previous scales in the same frame and the corresponding scale of the previous frames. 
Doing so gives us a powerful and efficient modeling ability to characterize arbitrary feature distributions. For instance, Mentzer {\it et al.}~\cite{mentzer2022vct} relied on a complicated prediction in a block-level autoregressive manner, which is inefficient. Instead, we perform a multi-stage conditional probability prediction, showing better performance and desiring less complexity. %Experimental evaluations demonstrate the efficacy of our proposed deep hierarchical predictive video coding. For example, 

Upon extensive evaluations using commonly used video sequences, our method outperforms well-known learned models using hybrid motion and residual coding and previous state-of-the-art method using latent probabilistic predictive coding. Extensive studies on the adaptation to various temporal patterns also reveal the generalization of our hierarchical predictive mechanism. In addition, our method also supports temporal progressive decoding, being the first learned progressive video coding method to our best knowledge. Therefore, it can handle packet losses induced by poor network connections to some extent. % Additional studies on the adaptation to various temporal patterns also reveal the generalization of our hierarchical predictive mechanism. 

Our contributions can be summarized as follows:

\begin{itemize}
    % \item We propose a hierarchical latent probabilistic predictive video coding method to explore the coarse-to-fine conditional distributions to better model the video frames to be encoded.
    \item {We propose a hierarchical probabilistic prediction model for video coding. Our model employs a collection of multiscale latent variables to represent the coarse-to-fine nature of video frames scale-wisely.}
    % \item With the design of hierarchical architecture and decoder side predictive fusion module, the proposed method achieves promising performance with light-weight computational complexity and fast decoding time.
    \item {We propose the spatial-temporal prediction and in-loop decoding fusion to enhance rate-distortion performance. Integrating these modules into the hierarchical architecture, the proposed method achieves better performance, lower memory consumption, and faster encoding/decoding than the previous best probabilistic predictive coding-based method~\cite{mentzer2022vct}.}
    \item Experiments demonstrate that our method is better generalized to various temporal patterns. Our model is also the first to support the functionality of progressive decoding. % Our model is also generalized to various temporal patterns with the proposed hierarchical predictive mechanism.
\end{itemize}

\section{Related Work} \label{sec:related_work}
We briefly review end-to-end learned video coding methods, including classical hybrid motion and residual coding and recently-emerged probabilistic predictive coding. We also theoretically explain the hierarchical VAE formalism as it provides the basis for our method.

% \begin{figure*}[htbp]
% \centering
% \subfloat[]{\includegraphics[scale=0.5]{figures/vae.pdf} \label{fig:vae}}
% \hspace{1.5cm}
% \subfloat[]{\includegraphics[scale=0.5]{figures/hyper_vae.pdf} \label{fig:hyper}}
% \hspace{1.5cm}
% \subfloat[]{\includegraphics[scale=0.5]{figures/resnet_vae.pdf} \label{fig:resvae}}
% \hspace{1.5cm}
% \subfloat[]{\includegraphics[scale=0.5]{figures/pred_vae.pdf}}
% \caption{{\bf Examples of VAE formalism.}}
% \end{figure*}

\def\ie{i.e.}
\def\eg{e.g.}
\def\kldiv{ D_\text{KL}}
\def\encoder{ f_\text{e} }
\def\decoder{ f_\text{d} }
\def\entmodel{ p(z) }

\subsection{Learned Video Coding} \label{sec:related_lvc}

\textbf{Data compression and variational autoencoders (VAEs):}
Let $x$ denote data (e.g., image or video) with an unknown distribution.
Traditional image/video coding belongs to the \textit{transform coding}, where one wants to find an encoder $\encoder$, a decoder $\decoder$, and an entropy model for the transform coefficients such that the rate-distortion cost is minimized:
\begin{equation}
\label{eq:related_transform_coding}
    \min H(\encoder(x)) + \lambda \cdot d(x, \decoder(\encoder(x))).
\end{equation}
Here, the first term is the (cross-) entropy of the compressed coefficients approximating the rate, $d$ is a distortion function, and $\lambda$ is the Lagrange multiplier that balances the rate and distortion tradeoff.
As studied in~\cite{balle2018variational, duan2023lossy}, transform coding can be equivalently considered as data distribution modeling using variational autoencoders (VAEs). Specifically, VAEs assume a model of data:
\begin{equation}
    p(x, z) = p(x \mid z) \cdot p(z),
\end{equation}
where $z$ is latent variables like transform coefficients.
In VAE, a \textit{prior} $p(z)$ describes the distribution of the latent variable, a decoder $p(x \mid z)$ maps latent-space elements to original data-space signal, and an \textit{approximate posterior} $q(z \mid x)$ (\ie, the encoder) encodes data into the latent space.
Letting $\hat{x} \sim p(x \mid z)$ denote the reconstruction, the objective can be written as~\cite{yang2022sandwich, duan2023lossy}
\begin{equation}
\label{eq:lossyvae}
    \min \kldiv(q(z \mid x) \parallel p(z)) + \lambda \cdot d(x, \hat{x}),
\end{equation}
and if the posterior $q(z \mid x)$ is deterministic and discrete (\eg, when quantization is applied to $z$), this VAE objective equals the rate-distortion optimization in Eq.~\eqref{eq:related_transform_coding}.
Such a connection has inspired many subsequent works to apply VAE-based probabilistic methods to compression task, such as~\cite{yang2020quantization, agustsson2020universally, yang2020improving, theis2022algorithms_comm_sapmles, ryder2022split_hierarchical, chen2022exploiting}.

% \textcolor{red}{(Need a description of the relationship between VAE and practical coding.)} Ball\'e {\it et al.}~\cite{balle2016end} bridged the gap between the VAE model and practical coding framework. Let $x$ be an image with unknown data distribution. In its original form, VAE models data by a joint distribution
% \begin{equation}
%     p(x, z) = p(x \mid z) \cdot p(z),
% \end{equation}
% where $z$ is an assumed \textit{latent variable}, $p(z)$ is the \textit{prior} distribution for $z$, and $p(x|z)$ is the conditional data likelihood of $x$ given $z$. VAE also defines an \textit{approximate posterior}, $q(z|x)$, to perform \textit{variational inference}. The three components of VAE, $p(z)$, $p(x|z)$, and $q(z|x)$, are all parameterized by neural networks, so from an autoencoder point of view, $q(z|x)$ can be interpreted as the encoder and $p(x|z)$ the decoder.

% Learned video coding methods mainly evolve from learned image coding by introducing the temporal priors to reduce the redundancy in interframe coding. At present, they can be generally categorized into the following two groups.
Learned video coding methods can be generally categorized into two groups: hybrid motion \& residual coding and probabilistic predictive coding.

\textbf{Hybrid Motion \& Residual Coding} refers to the classical coding framework with motion and residual processing. Lu et al.~\cite{lu2019dvc} first proposed to use two similar VAE-based networks to code the motion and residuals, respectively, which was then enhanced with better motion alignment in \cite{lu2020end,liu2020neural}. Then, Hu et al.~\cite{hu2021fvc} migrated the motion alignment to the feature domain and achieved better compression performance. Recently, by converting residual coding to conditional coding of aligned features, Li et al.~\cite{li2021deep} took the learned video coding to a new level of performance. Subsequently, by further integrating multi-scale aligned feature fusion, post-processing, and bitrate allocation, learned video coding algorithms achieved unprecedented compression efficiency, surpassing the latest VVC~\cite{li2022hybrid}.

\textbf{Probabilistic Predictive Coding} is an emerging video coding method. Liu et al.~\cite{liu2020conditional} relied on stacked convolutions for latent distribution prediction, while VCT~\cite{mentzer2022vct} adopted Transformer for the same purpose. Both works perform temporally-conditional distribution prediction only using single-scale latent variables  (i.e., $1/16$ of the original resolution), which greatly constrains the accuracy of probability estimation and leads to sub-optimal predictive performance. Therefore, in this paper, we propose a hierarchical probabilistic predictive coding method, which substantially improves the accuracy and efficiency of temporal prediction by characterizing multiscale latent features for conditional probability estimation in a coarse-to-fine approach.

\subsection{Hierarchical VAEs}
\label{sec:related_hvae}
To improve the flexibility and expressiveness of single-sale VAE, hierarchical VAEs~\cite{kingma2016improved, child2020very, vahdat2020nvae} employ multiscale latent variables, denoted by $Z = \{z^1, ..., z^L\}$. Accordingly, the latent priors can be factorized as:
\begin{equation}
    p(Z) = \prod_{l=1}^L p(z^l \mid Z^{<l}),
\end{equation}
where $L$ is the total number of hierarchical scales, and $Z^{<l}$ denotes $\{z^1, ..., z^{l-1}\}$. Typically, $z^1$ has the smallest dimension, while $z^L$ is with the largest dimension. Such a dimensional {refinement} from a lower scale to a higher one improves the flexibility of VAEs and effectively captures the coarse-to-fine characteristics of images.

Among popular hierarchical VAE architectures, ResNet VAE~\cite{kingma2016improved} provides the most promising performance in terms of image modeling. Different from the Hyperprior VAE used in \cite{balle2018variational}, each latent variable $z^l$ of ResNet VAE is conditioned on all $Z^{<l}$ and its encoding is bi-directional with dependent on both $x$ and $Z^{<l}$. This might explain the fact that ResNet VAE can be scaled up to more than 70 layers~\cite{child2020very}. The loss function for training ResNet VAEs can be extended from Eq.~\eqref{eq:lossyvae} for supervising multiscale latents:
\begin{equation} 
\label{eq:lossyhvae}
    \min \sum_{i=1}^L \kldiv(q^l \parallel \, p^l)
    +
    \lambda \cdot d(x, \hat{x}),
\end{equation}
where $q^l$ and $p^l$ are shorthand notations for the posterior and prior for the $l$-th scale latent variable, i.e.,
\begin{equation}
\label{eq:qarv_related_notation}
\begin{aligned}
    q^l  = q(z^l \mid x,Z^{<l}),\mbox{~~and~~} p^l  = p(z^l \mid Z^{<l}).
\end{aligned}
\end{equation}

Our proposed method is developed based on the ResNet VAE structure by further introducing the temporal priors in addition to the deliberate design for practical compression.

% We iteratively sample $z^l$ from $q^l$ for $l = 1, 2, ..., L$, where at each step $q^l$ is conditioned on the previously sampled value of $Z^{<l}$. During inference, $z^l$ is also conditionally dependent on $x$ in addition to $Z^{<l}$.

% As we have described in Sec.~\ref{sec:related_work}, VAE has shown good performance in learned image/video compression. To further improve the flexibility and expressiveness of VAE, hierarchical VAEs~\cite{kingma2016improved,child2020very} have been proposed, which employ $N$ latent variables $Z = \{Z_1, Z_2, ..., Z_N\}$ as
% \begin{equation}
%     p_{Z^{1:L}} = p_{Z^L|Z^{<L}} \, \cdots \ p_{Z^3|Z^2,Z^1} \cdot p_{Z^2|Z^1} \cdot p_{Z^1},
% \end{equation}

% Duan {\it et al.}~\cite{duan2023lossy} proposed a lossy image compression method using a hierarchy latent model. 
\section{Preliminary: Predictive Video Coding}
% \section{preliminaries}
% We first have an illustration of the probabilistic predictive video coding. Given a random video sequence $X = \{ x_1, ..., x_T\}$, which contains $T$ frames to be encoded. Our goal is to learn a video coding model to ensure favorable reconstructed results at the least bitrate consumption.

Suppose a video sequence $X = \{ x_1, ..., x_T\}$ that contains $T$ frames for encoding. As a convention for predictive coding methodologies~\cite{liu2020conditional,mentzer2022vct}, an analysis encoder followed by quantization is first applied to transform each input frame ${x}_t$ into the discrete latent representation ${z}_t$ with reduced resolution. A symmetrical decoder is then used to recover the reconstruction $\hat{{x}}_t$ from ${z}_t$. Having the probability mass function (PMF) $p(z_t)$ to estimate the true distribution of symbols in $z_t$, we can get the bits desired for transmission by entropy coding. The main idea of the probabilistic predictive coding is to parameterize the $p(z_t)$ as a conditional distribution
\begin{align}
\label{eq:latent_condition}
p(z_t) = p(z_t \mid Z_{<t}),
\end{align}
where $Z_{<t}$ is a set of latent features preceding time $t$. By exploiting the temporal redundancy across frames using an efficient prediction network, one can obtain more accurate probability estimation for the current frame to reduce cross-entropy and thus maximize coding efficiency. To this end, VCT~\cite{mentzer2022vct} %follows the latent prediction formalism and 
introduces a Transformer model in a block-level autoregressive manner to model $p(z^i_t \mid z^{<i}_t, Z_{<t})$ for jointly appreciating spatial and temporal correlation. Here, $z^i_t$ corresponds to a pixel at position $i$ within a predefined block in the latent space of the current frame, and $z_t^{<i}$s are autoregressive neighbors previously processed.

Although decent compression performance is achieved, VCT only performs conditional probability prediction using single-scale latent variables, ignoring the multiscale characteristics in both spatial and temporal domains.
Therefore, the prediction is sub-optimal, and the complexity of the prediction network is usually unaffordable.
% \zhihao{
% Despite its reasonable compression performance, VCT performs conditional probability prediction on single-scale latent variables, ignoring the multiscale characteristics of videos in both spatial and temporal domains.
% Therefore, this paper proposes ...
% }

% We train $E$, $D$ using standard neural image compression techniques to be lossy transforms reaching nearly any desired distortion $d({\boldsymbol x}_t, \hat{{\boldsymbol x}}_t)$ by varying how large the range of each element in ${\boldsymbol z}_t$ is. Given a probability mass function (PMF) $P$ estimating the true distribution $Q$ of symbols in ${\boldsymbol z}_t$, we can use entropy coding (EC) to transmit ${\boldsymbol z}_t$ with $H \cdot W \cdot C \cdot E_{z \sim Q({\boldsymbol z})}[-log_{2}p(z)]$ bits. The main idea of the predictive coding is to parameterize $p$ as a conditional distribution.

\section{Proposed Method}

\begin{figure*}[htbp]
\centering
\subfloat[]{\includegraphics[width=0.3\linewidth]{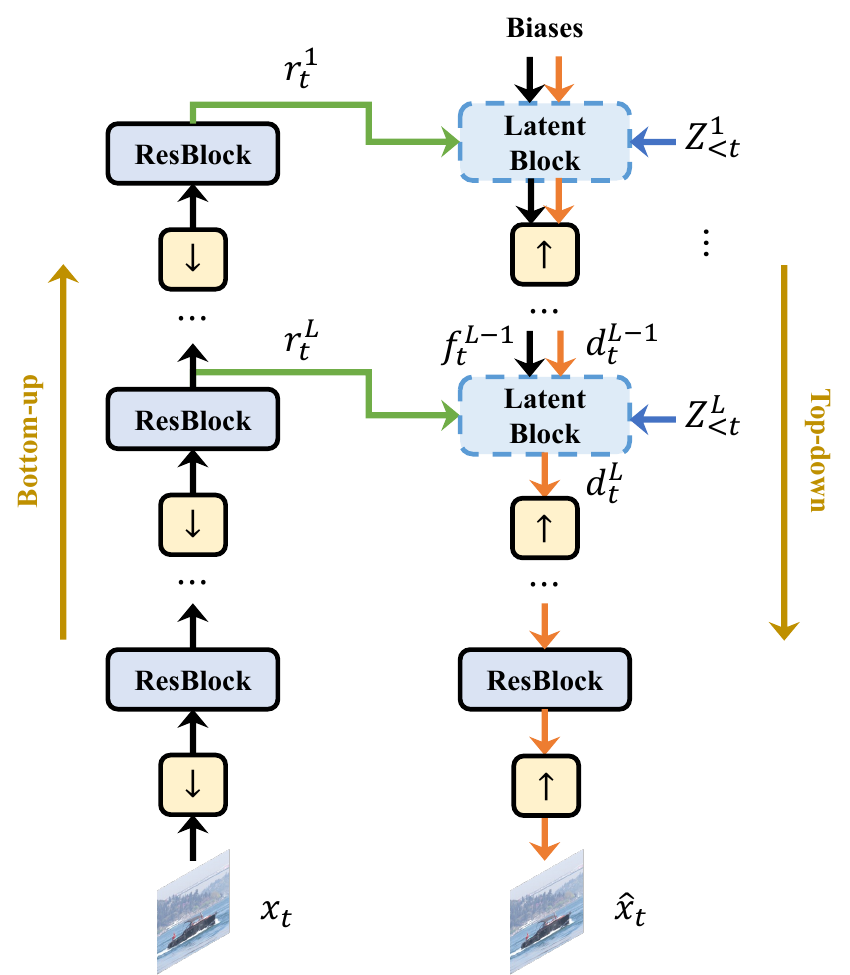} \label{fig:network}}
\hspace{0.24cm}
\subfloat[]{\includegraphics[width=0.35\linewidth]{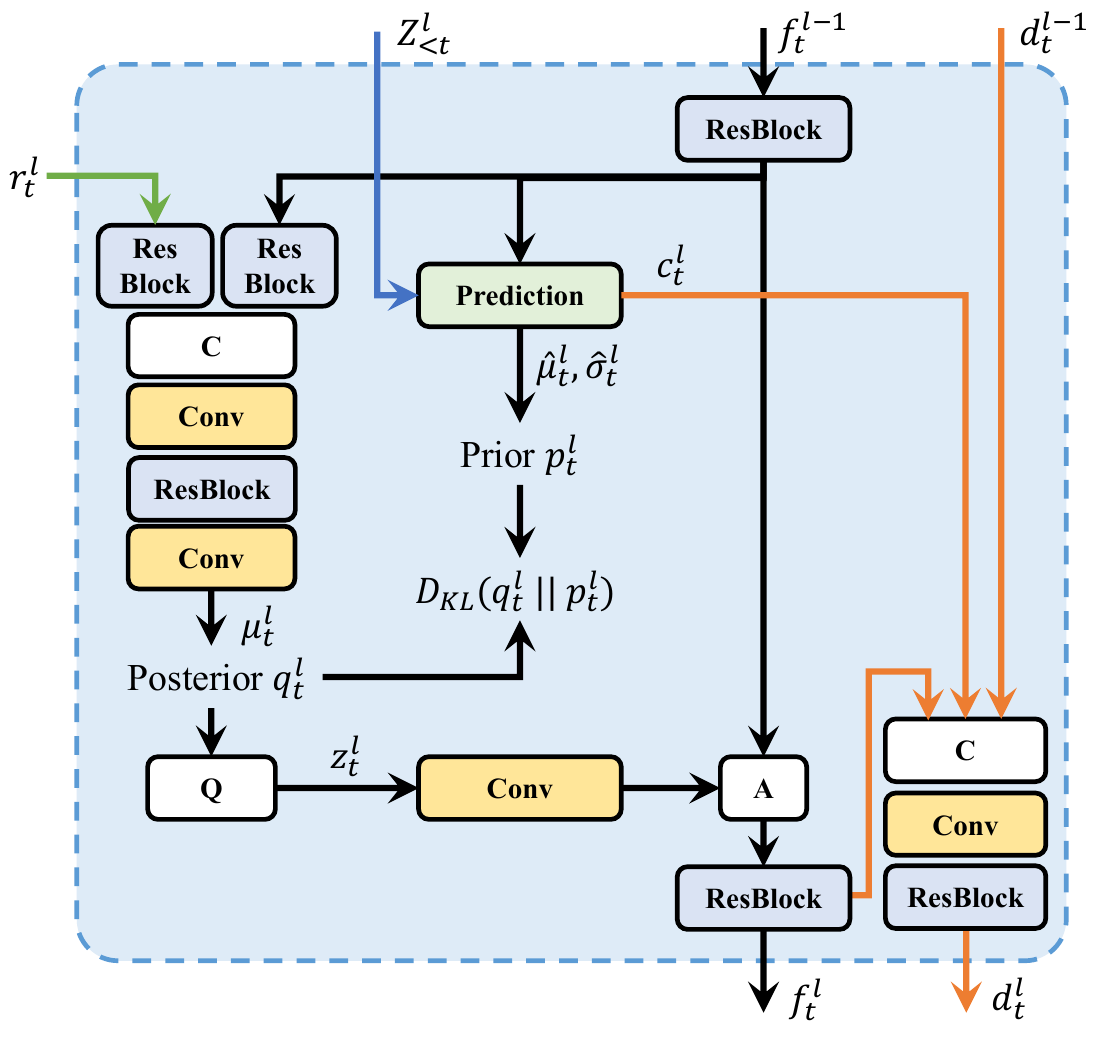}\label{fig:latent_block}}
\hspace{0.48cm}
\subfloat[]{\includegraphics[width=0.22\linewidth]{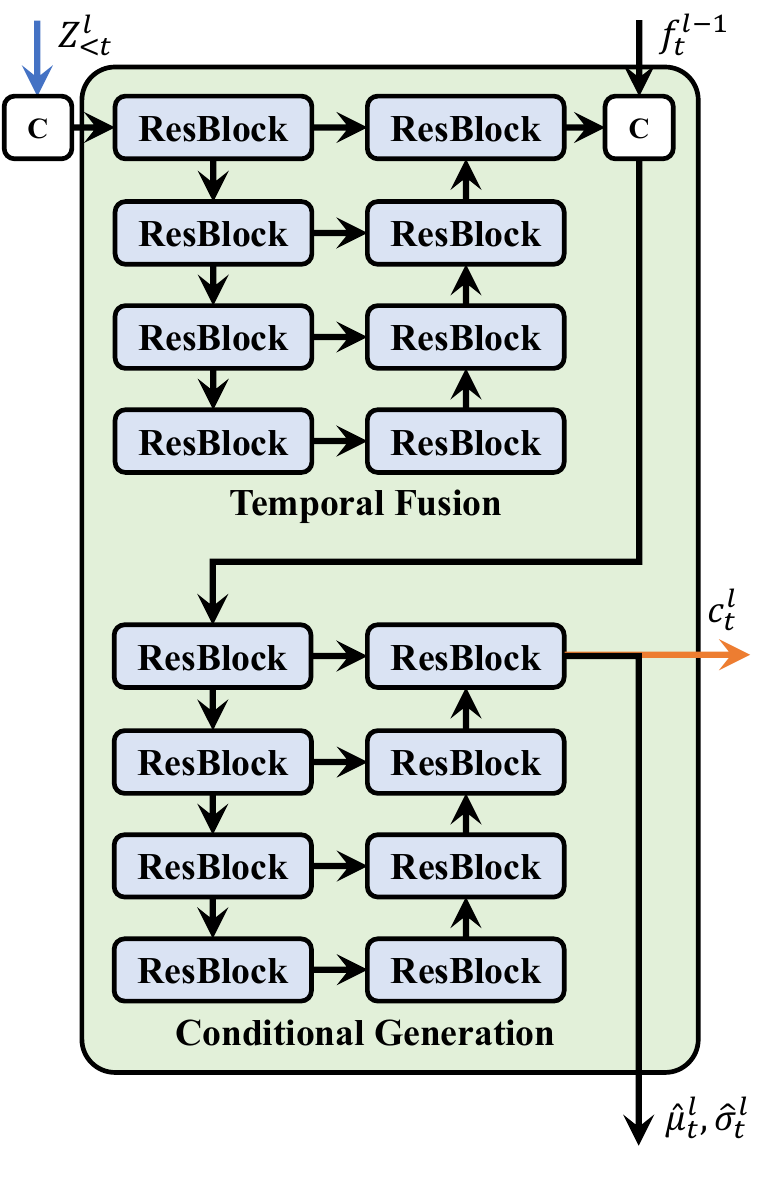}\label{fig:prediction_network}}
\caption{(a) {\bf Overall Architecture}, (b) {\bf Latent Block}, and (c) {\bf Spatial-Temporal Prediction Module} of our proposed DHVC. $\downarrow$ and $\uparrow$ are respective downscaling and upscaling operations. ``C'' represents concatenation, ``A'' represents addition, and ``Q'' represents quantization. The convolutional layer ``Conv'' is used for feature re-dimension.}
\vspace{-0.2cm}
\end{figure*}

% \begin{figure}[htbp]
% \centering
% \includegraphics[scale=0.4]{figures/latent_block.pdf}
% \caption{{\bf Detailed Architecture of the Latent Block.}}
% \label{fig:latent_block}
% \end{figure}

% \begin{figure}[htbp]
% \centering
% \includegraphics[scale=0.35]{figures/prediction.pdf}
% \caption{{\bf Prediction Network.}}
% \label{fig:prediction_network}
% \end{figure}

% In this section, we first overview the network architecture of our proposed model. Then, we detail the hierarchical probabilistic predictive mechanism and the construction of the posteriors, priors, and decoder.

\subsection{Network Architecture Overview}
% We deliberately employ ResNet VAEs units in our work with network architecture briefed below first.

%\textbf{Overall Architecture:}
Figure~\ref{fig:network} depicts the overall framework of our method, which consists of a \textit{bottom-up} path and a \textit{top-down} path.
Given an input frame $x_t$, the {bottom-up path} produces a set of features $R_t=\{r_t^1, ..., r_t^L\}$ at respective $1/64$, $1/32$, $1/16$, and $1/8$ resolutions of the original input through scale-wise downsampling and information aggregation/embedding using residual blocks (ResBlocks). Different from \cite{chen2022exploiting}, the $r_t^l$ at each scale is dependent on both feature extracted from $x_t$ and decoded feature from lower scale, which turns our method into conditional residual coding to minimize the bitrate consumption.
These residual features $R_t$ are subsequently sent to the {top-down path} for hierarchical probabilistic modeling.
% which are subsequently sent to the top-down path for (approximate) inference.

The top-down path starts from two learnable constant biases and then encodes a sequence of latent variables $Z_t=\{z_t^1, ..., z_t^L\}$ (in the Latent Blocks) to produce respective prior feature $f_t^l$ and reconstructive feature $d_t^l$ scale-by-scale. In the end, the $\hat{x}_t$ is reconstructed by passing the last reconstructive representation $d_t^L$ into multiple upsampling and ResBlock layers. The down-sampling operations $\downarrow$ are implemented by strided convolution, and the up-sampling operations $\uparrow$ are implemented by $1\times1$ convolution followed by pixel shuffle. The ConvNeXt~\cite{liu2022convnet} units are adopted in ResBlocks. More details can be found in supplementary materials.

% Note that a single model is used for both intraframe and interframe coding. The temporal priors are set using learnable constant biases for intraframe coding, while they are generated recurrently from corresponding temporal references for interframe coding. 

%features used for the intra frame are replaced by a set of .

%, the compression process starts with a learnable constant bias following into a sequence of latent blocks (shown in Fig.~\ref{fig:latent_block}), 

%\textbf{Network Component:}

%are used for information encoding, i.e., $1/64$, $1/32$, $1/16$, and $1/8$, and the number of latent blocks at each resolution is 1, 2, 1, and 1 respectively. 

\subsection{Predictive Coding Modules}
\label{sec:method_pcm}
{
We now detail the architecture of the Latent Block (Fig.~\ref{fig:latent_block}), which is critical to the effectiveness of our approach. Like in ResNet VAEs, each Latent Block adds ``information", carried by the latent variable $z_t^l$, into the top-down path features.
We substantially extend it by introducing (1) a Spatial-Temporal Prediction Module for predictive coding and (2) an In-loop Decoding Fusion Module to improve coding performance, which is described below one-by-one.
}

\textbf{Spatial-Temporal Prediction Module} (Fig.~\ref{fig:prediction_network}):
% The detailed prediction module is shown in Fig.~\ref{fig:prediction_network}.
{
To predict $z_t^t$ at $l$-th scale, we combine the same-scale temporal priors $Z_{<t}^l$ with spatial prior $f_t^{l-1}$ from previous scales to produce the prior distribution parameters.
% We use the two previous latents $z_{t-1}^l$ and $z_{t-2}^l$ to make up of $Z_{<t}^l$.
}
% Two previous latents $z_{t-1}^l$ and $z_{t-2}^l$ make up of the temporal priors $Z_{<t}^l$ in this work.
We begin with the Temporal Fusion by passing the temporal priors $Z_{<t}^l$ into stacked ResBlocks, with skip connections at each level. Then, the spatial prior feature $f_t^{l-1}$ is concatenated with the fused temporal information for subsequent Conditional Generation to get the contextual feature $c_t^l$ and the prior distribution parameters, i.e., the mean $\hat{\mu}^l_t$ and scale $\hat{\sigma}_t^l$.
% The network of Conditional Prediction shares the same architecture as the Temporal Fusion except for the output. At the end of the Temporal Prediction Module, we will get the predicted contextual feature $c_t^l$ and the prior distribution parameters, i.e., the mean $\hat{\mu}^l_t$ and scale $\hat{\sigma}_t^l$.
% Note that the prediction performance can be further improved by more sophisticated prediction algorithms or spatial and/or channel autoregressive model.

% \textbf{Decoder:}
\textbf{In-loop Decoding Fusion Module} (on the right of Fig.~\ref{fig:latent_block}):
Two distinct features are generated during the decoding process: the prior feature $f_t^l$ utilized as the spatial prior for subsequent scale, and the reconstructive feature $d_t^l$ for the eventual result reconstruction. In our specific implementation, we concatenate the previously decoded feature $d_t^{l-1}$, and the contextual feature $c_t^l$, along with the $f_t^l$ to generate the fused results $d_t^l$. This design represents a notable departure from the original ResNet VAE framework, which employs a single top-down path feature for both the prior and reconstruction purposes. Through the method in our study, the $f_t^l$ solely handles conditional distribution modeling, whereas the $d_t^l$ is responsible for the reconstruction aspect. By leveraging the dependable contextual feature $c_t^l$, we achieve a desirable decoded $d_t^l$ while conserving bitrate consumption effectively.

{
Note that our framework requires only a single model for intra and inter frame coding. For intra-coding, the temporal priors $Z_{<t}^l$ each scale are set using learnable constant biases, while for inter coding, we use the two previous latents $z_{t-1}^l$ and $z_{t-2}^l$ to make up of $Z_{<t}^l$.
}

% \subsection{Hierarchical Probabilistic Modeling}
\subsection{Probabilistic Model and Loss Function}
\label{sec:method_hierarchical_pm}

% where each latent block adds ``information", carried by the latent variables $z_t^l$, into the prior feature $f_t^l$ and decoded feature $d_t^l$ at level $l$. The additional temporal latent features $Z_{<t}^l$ are also introduced for probabilistic predictive modeling using the prediction network in Fig.~\ref{fig:prediction_network}.

% \zhihao{
% Each latent block adds ``information", carried by the latent variables $z_t^l$, into the prior feature $f_t^l$ and decoded feature $d_t^l$ at level $l$.
% Our framework extends the ResNet VAEs~\cite{kingma2016iafvae} to predictive video coding by conditioning $z_t^l$ on temporal latent variables $Z_{<t}^l$.
% }

% \zhihao{
% This section presents the proposed hierarchical probabilistic model for predictive video coding.
% We start with the probabilistic framework (posterior and prior distributions), and we then detail our neural network implementation of it (temporal prediction and decoder-side fusion modules).
% }

% As mentioned in \ref{sec:related_lvc}, to support the practical lossy compression using the feasible entropy coding algorithms, we follow the previous work~\cite{balle2018variational,duan2023lossy} and apply the quantization-aware training by using uniform posteriors.
{
With the specific neural network modules, our framework effectively extends hierarchical VAEs to predictive video coding.
% As mentioned in \ref{sec:related_hvae}, our probabilistic model is based on (but substantially extends) ResNet VAEs.
To support practical lossy compression using feasible entropy coding algorithms, we follow previous works~\cite{balle2018variational,duan2023lossy} and apply quantization-aware training using uniform posteriors.
}
Specifically, we adopt a hybrid quantization strategy at training time to simulate the quantization error. The additive uniform noise is applied in terms of rate estimation, while the straight-through rounding operation is used for reconstruction. We use uniform quantization at test time. For the prior, we use the Gaussian distribution convolved with uniform distribution, which is flexible to match the posterior.

\textbf{Posteriors:}
The (approximate) posterior for the $l$-th latent variable, $z_t^l$, is defined to be an uniform distribution:
\begin{equation}
q(z_t^l \mid x_t, Z_t^{<l}) = U(\mu_t^l - \frac{1}{2}, \mu_t^l + \frac{1}{2}),
\end{equation}
where $\mu_t^l$ is the output of the posterior branch in the latent block (see Fig.~\ref{fig:latent_block}) by merging the embedded feature $r_t^l$ and prior feature $f_t^{l-1}$ from the previous level. The discrete $z_t^l$ depends on the frame $x_t$ as well as previous level latent variables $Z_t^{<l}$.

\textbf{Priors:}
{
Our framework extends ResNet VAE to predictive video coding by conditioning the prior distributions for $z_t^l$ on temporal latent variables $Z_{<t}^l$.
}
At each timestep, considering $L$ levels of latent variables $Z_t = \{z_t^1, ..., z_t^L\}$, the latent conditional distribution can be factorized as
\begin{align}
    p(Z_t \mid Z_{<t}) = \prod_{l=1}^Lp(z_t^l \mid Z_t^{<l}, Z_{<t}^l),
\end{align}
Then, the prior distribution for each $z_t^l$ is defined as a Gaussian convolved with a uniform distribution:
\begin{equation}
\label{eq:method_prior}
\begin{aligned}
p(z_t^l \mid Z_t^{<l}, Z_{<t}^l) = \mathcal{N}(\hat{\mu}_t^l, {\hat{\sigma}_t^l}\textsuperscript{\textsuperscript{2}}) * U(- \frac{1}{2}, \frac{1}{2})
\end{aligned}
\end{equation}
where $\mathcal{N}(\hat{\mu}_t^l, {\hat{\sigma}_t^l}\textsuperscript{\textsuperscript{2}})$ denotes the Gaussian probability density function. The mean $\hat{\mu}^l_t$ and scale $\hat{\sigma}_t^l$ are predicted by the prior branch in the latent block. Note that the prior mean $\hat{\mu}_t^l$ and scale $\hat{\sigma}_t^l$ are dependent on both the latent variables from previous time steps $Z_{<t}^l$ for that level and on the latent variables of the previous levels at the current timestep $Z_t^{<l}$.

\begin{figure*}[htbp]
\centering
\subfloat[]{\includegraphics[width=0.25\linewidth]{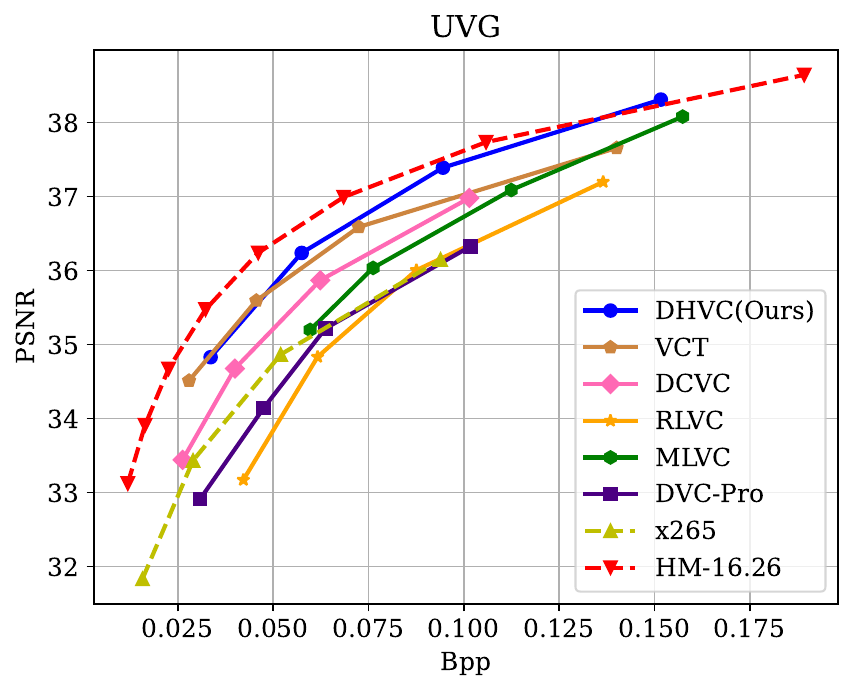}}
\hspace{0.5cm}
\subfloat[]{\includegraphics[width=0.25\linewidth]{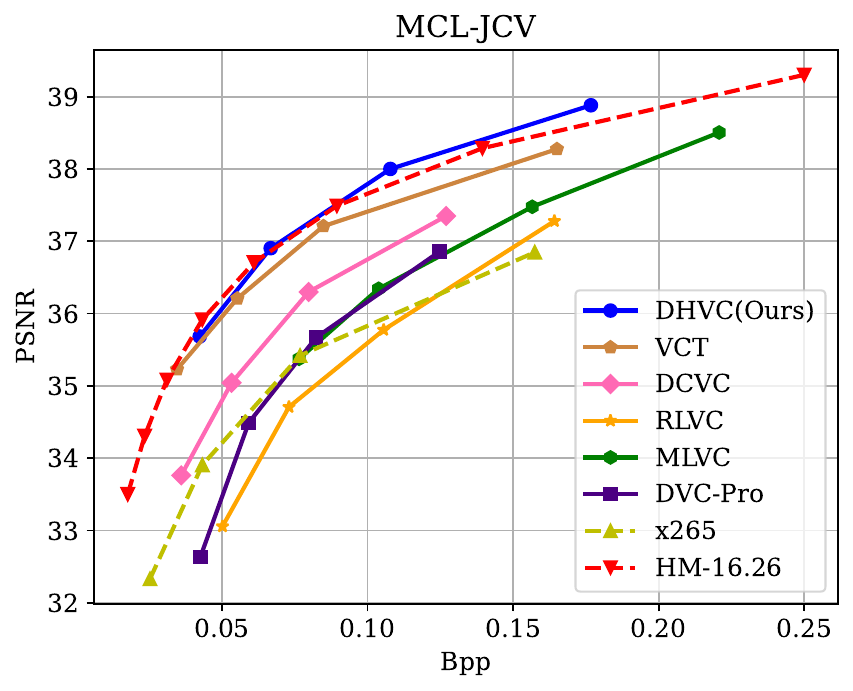}}
\hspace{0.5cm}
\subfloat[]{\includegraphics[width=0.25\linewidth]{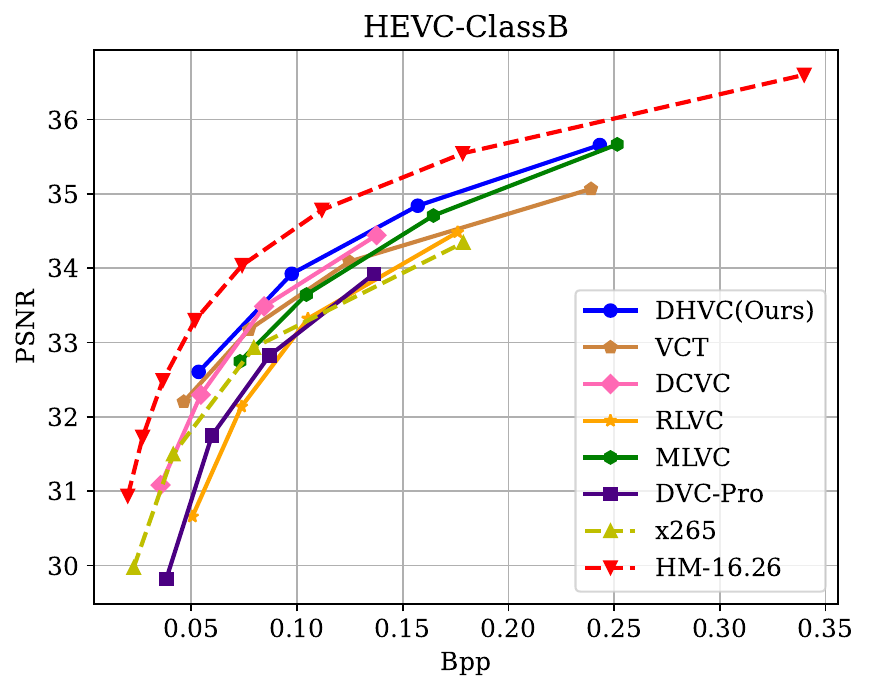}} \\
\subfloat[]{\includegraphics[width=0.25\linewidth]{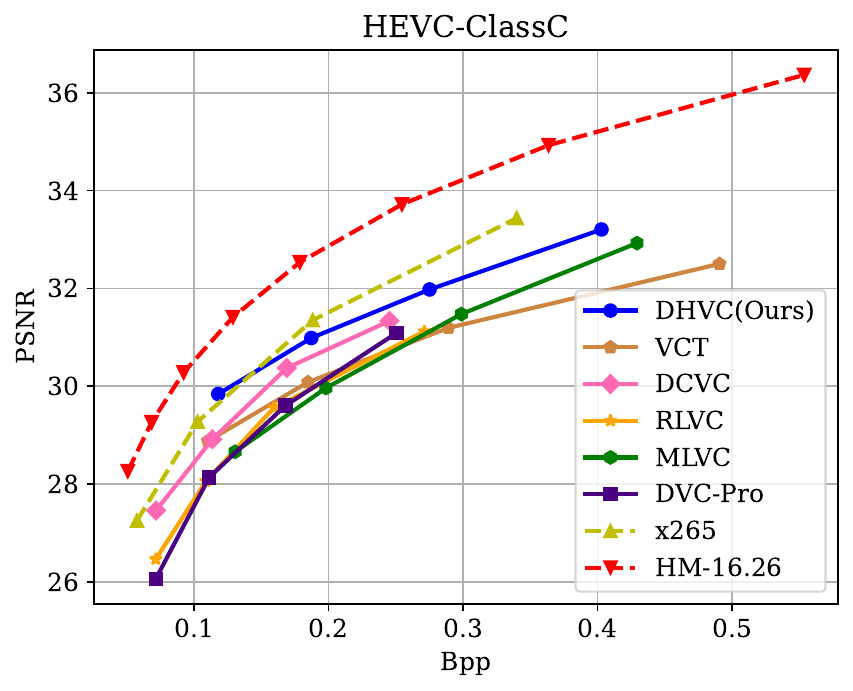}}
\hspace{0.5cm}
\subfloat[]{\includegraphics[width=0.25\linewidth]{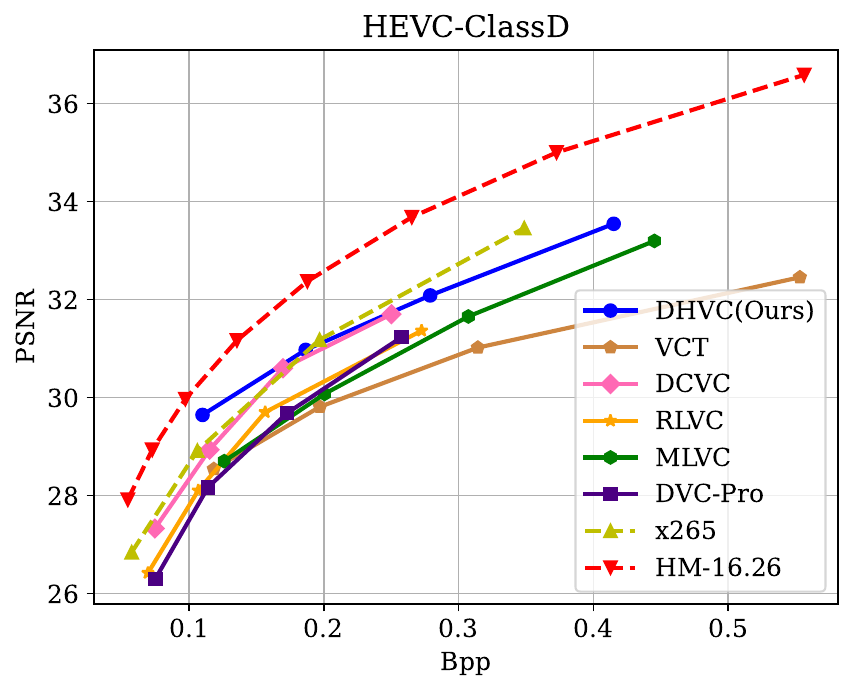}}
\hspace{0.5cm}
\subfloat[]{\includegraphics[width=0.25\linewidth]{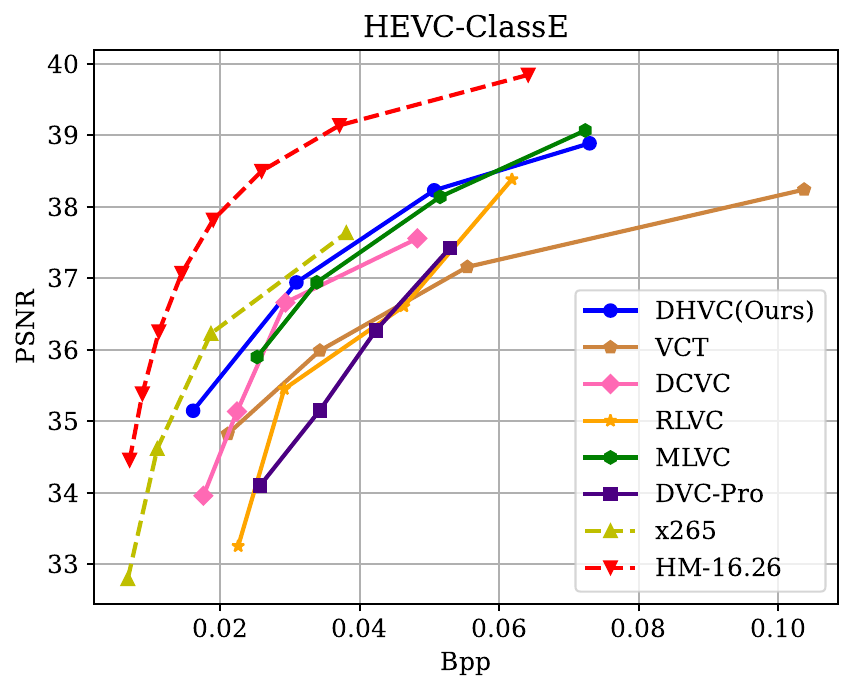}}
\caption{{\bf Compression efficiency} comparison using rate-distortion (R-D) curves: learned video coding models are trained using MSE loss; both the H.265/HEVC reference model targeting for performance study, HM-16.26, and open-source model targeting for real-time application, x265, are used.}
\label{fig:rd_curve}
\vspace{-0.2cm}
\end{figure*}

\textbf{Training Objective:}
Typically, the hybrid motion and residual coding methods require multi-stage or simultaneous optimization of the optical flow, motion coding, and residual coding networks during the training phase. Differently, the training process of our model is as easy as optimizing a lossy image coding. The loss function $\mathcal{L}$ is extended from Eq.~\eqref{eq:lossyhvae} with the inclusion of temporal dependency:
\begin{align}
    \mathcal{L} 
   %  &= min
   % \sum_{l=1}^L \kldiv (q^l_t \parallel \, p^l_t) + \lambda \cdot d(x_t, \hat{x}_t)
   %  \\
    = \min
    \sum_{l=1}^L -\log_2 {p(z_t^l \mid Z_t^{<l}, Z_{<t}^l)}
    +
    \lambda \cdot d(x_t, \hat{x}_t), \!
\end{align}
The first term consists of the rate for all latent variables. The second term corresponds to the reconstruction distortion, which is commonly chosen to be the mean squared error (MSE) or MS-SSIM~\cite{wang2003multiscale} loss for videos. The multiplier $\lambda$, which trades off rate and distortion, is pre-determined and fixed throughout training. At test time, the $\lambda$ is the same for both intra and inter frame, and the actual bitrates are determined by the estimation accuracy of conditional probabilistic modeling.

% \textbf{Latent Blocks:}
% The latent block modules are used for feature coding, as can be seen in Fig.~\ref{fig:latent_block}. The left branch is used for feature encoding. $R_t^l$ denotes the feature extracted from the input frame and $F_t^{l-1}$ is the prior features from the proceeding latent block. We fuse $R_t^l$ and $F_t^{l-1}$ through several convolutional blocks and get the latent representation $z_t^l$ after quantization. The prior feature $F_t^{l-1}$ is input into the prediction network with the proceeding features $F_{<t}^{l}$ for conditional probabilistic distribution prediction. Furthermore, the prior feature $F_t^{l-1}$ will be also added into the convolved $z_t^l$ to generate the $F_t^{l}$ for the succeeding latent block. In addition, the new prior feature $F_t^{l}$, the predictive feature $C_t^{l}$, and the proceeding decoded feature $D_t^{l-1}$ will be concatenated to obtain the current decoded results $D_t^{l}$.

% \begin{figure}[htbp]
% \centering
% \includegraphics[scale=0.5]{figures/pipeline.pdf}
% \caption{{\bf Pipeline.}}
% \label{fig:pipeline}
% \end{figure}

% \textbf{Prediction:}
\section{Experimental Results}
\subsection{Implementation Settings}
% We clarify the training and test datasets as well as the training details below:

\textbf{Datasets:}
We use the popular Vimeo-90K~\cite{xue2019video} dataset to train our model, which consists of 64,612 video samples. % with a fixed resolution of $448 \times 256$. 
Training batches comprise sequential frames that are randomly cropped to the size of $256 \times 256$. Commonly used test datasets, i.e., the UVG~\cite{mercat2020uvg}, MCL-JCV~\cite{wang2016mcl}, and HEVC Class B, C, D, and E~\cite{bossen2013common}, are used for evaluation. They cover various scene variations, and resolutions are available from $416\times240$ to $1920\times1080$. Test sequences in YUV420 format are pre-processed following the suggestions in \cite{9941493} to generate RGB frames as the input of learned models. The first 96 frames of each video are used for evaluation, and the group of pictures (GOP) is set at 32. {These settings are the same as in other learned video coding methods.}

\textbf{Training details:}
We progressively train our model for fast convergence. First, the model is trained to encode a single frame independently for 2M iterations by {setting the temporal prior at each scale level as a learnable bias. Then, we train the aforementioned model for 500K steps using three successive frames, with temporal priors hierarchically generated from previously-decoded frames. In the end, another 100K steps are applied to fine-tune the model using five successive frames, for which it better captures long-term temporal dependence~\cite{liu2020neural}. We set $\lambda$ from \{256, 512, 1024, 2048\} and \{4, 8, 16, 32\} for respective MSE and MS-SSIM optimized models to cover wide rate ranges. Adam~\cite{kingma2014adam} is the optimizer with the learning rate at $10^{-4}$. Our model is trained using two Nvidia RTX 3090, and the batch size is fixed at eight.

%introduce triplets of adjacent frames to have temporal probabilistic prediction with the previous decoded hierarchical representations for 500K steps.}} 

\subsection{Evaluation}
All the evaluation experiments are performed under the low-delay configuration. We choose x265~\footnote{\url{https://www.videolan.org/developers/x265.html}} and HM-16.26~\footnote{\url{https://hevc.hhi.fraunhofer.de}} as the benchmarks of traditional video codecs. Both of them use the default configuration.}
The detailed codec settings can be found in supplementary materials. For learned video coding methods, we compare with the representative algorithms using hybrid motion \& residual coding method, i.e., DVC-Pro~\cite{lu2020end}, MLVC~\cite{lin2020m}, RLVC~\cite{yang2020learning}, and DCVC~\cite{li2021deep}, and the best-performing probabilistic predictive coding model VCT~\cite{mentzer2022vct}. {Recently, a set of codec optimization tools, like feature fusion, post-processing, bitrate allocation, etc., are integrated into the representative frameworks to further improve the compression performance~\cite{li2022hybrid}. In this work, we currently focus on comparisons regarding frameworks themselves, and choose the best-in-class methods without further augmentation. Integrating these optimization mechanisms into our method is an interesting topic for future study.}

% For comparison, we choose the traditional video codecs, i.e., x265 and HM 16.26 with low-delay profile. It should be noted that the HM 16.26 configuration is using more than one reference frames to achieve the optimal compression performance, which can be served as a strong comparison baseline. For learned video coding methods, we choose to use the representative motion compensated work, DVC-Pro~\cite{lu2020end}, MLVC, RLVC, and DCVC~\cite{li2021deep}, and the SOTA probabilistic predictive coding method VCT~\cite{mentzer2022vct} for comparison. The performance of these motion compensated work can be easily found in their paper or open resourced website. While as per VCT, since there is no pre-trained resources for evaluation, we retrain the model using the same dataset as ours for fair comparison.

% \begin{table}[t]
% \footnotesize
% \centering
% \begin{tabular}{ccccccc}
% \toprule
% Method & UVG & MCL-JCV & ClassB & ClassC & ClassD & ClassE \\ 
% \midrule
% HM &  &  & &&& \\
% DCVC &  &  & &&& \\
% VCT &  &  & &&& \\
% DHVC (Ours) &  &  & &&& \\
% \bottomrule
% \end{tabular}
% \caption{BD-rate over x265}
% \end{table}

\textbf{Performance:} 
Figure~\ref{fig:rd_curve} depicts the R-D (rate-distortion) curves for testing various methods across various datasets. Our model, DHVC, leads all the learned methods regardless of testing videos, revealing the generalization of our method. It also performs much better than x265 on popular datasets with 1080p resolution (e.g., UVG and MCL-JCV) and even better than HM-16.26 on the MCL-JCV, suggesting the encouraging potential of such hierarchical predictive coding.

Though our method still performs the best, we must admit that it yet remains a noticeable performance gap between the learned codecs and traditional ones on low-resolution videos in HEVC Class C, D, and E. {{We believe this mainly owes to the conditional probability estimation or motion/residual coding upon latent variable with much lower resolution downscaled from the original input. For a low-resolution input, its local block contains more complicated texture patterns than that in a same-size block of a high-resolution video. Thus, the downsampling-induced information loss is more critical, potentially leading to inaccurate correlation characterization and subsequent compression. Such a hypothesis is well justified when comparing the performance between the VCT and our method. VCT shows a great performance drop for those low-resolution datasets, even worse than the earliest learned method DVC-Pro in middle and high bitrate ranges. This is because VCT conducts the conditional probability estimation upon single-scale latent features at 1/16 resolution of the original input. Instead, our method performs conditional probability estimation using multi-scale latent variables at respective 1/64, 1/32, 1/16, and 1/8 resolutions. Such a hierarchical mechanism greatly improves performance by thoroughly exploiting the coarse-to-fine natural characteristics of video frames.}}

% Concerning the performance loss of learned codecs to the HM-16.26 on HEVC Class B videos at 1080p resolution, it is because samples in Class B present more intense motions than those in MCL-JCV and UVG. As seen, dedicated variable-size motion estimation/compensation in traditional codecs can better handle these cases. Additional codec optimizations studied in~\cite{li2022hybrid} can be borrowed to enhance the proposed DHVC for even better performance.

%For example, learned codecs typically apply multi-stage resolution downsampling to derive latent variables for compression, which would inevitably introduce information loss iownsampling would introduceeffective information in low-resolution videos. }} 

%It is worth noting that similarly as a predictive coding method, the performance of VCT is mediocre. 
% This owes to the fact that VCT conducts conditional probability estimation upon even lower resolution latent features, e.g., 1/16 of the original video, for which resolution down-sampling 

%An important reason for this is that at such low input resolution, the probabilistic predictive coding such as VCT is usually done on latent features with even lower resolution. This seriously degrades the expressiveness of the frame probability distribution and thus the accuracy of the temporal prediction. Our proposed method greatly improves the accuracy of inter-frame conditional probability modeling due to the introduction of hierarchical feature coding, which exploits the coarse-to-fine natural characteristics of images.

Due to the space limitation, results for learned models trained using MS-SSIM loss can be found in the supplemental materials, which show a clear advantage of our method over both the traditional and learned video codecs.

\textbf{Complexity:} 
%We have a complexity analysis of several learned video coding methods to illustrate the efficiency of our proposed method. 
{Evaluation results with 1080p videos are listed in Table~\ref{tbl:complexity}. Except for the model size, our method shows clear advantages for other metrics, reporting the least requirements of respective kMACs per pixel, peak memory consumption, encoding, and decoding time. This also suggests that the model size is not closely related to the computational complexity of running codecs in practice. The sizeable parameters used in our method are mainly attributed to using basic ConvNeXt units to form the ResBlocks and Latent Blocks (see Fig.~\ref{fig:network}).}

% We can greatly reduce the model parameters using simple ResNet units~\cite{he2016deep}, but striking a justified tradeoff between performance and complexity requires substantial efforts in the future.}

\begin{table}[t]
% \scriptsize
\centering
\begin{scriptsize}
\begin{tabular}{cccccc}
\toprule
Method & Params (M) & kMACs/pix & PeakMem (G) & ET (s) & DT (s) \\ 
\midrule
% DVC-Pro & \textbf{37.82+} & 855.06 & 245 / 113 \\
DCVC & \textbf{7.58} & 1126.40 & 11.71 & 13.78 & 46.48 \\
VCT & 187.82 & 3042.20 & 10.51 & 1.64 & 1.58 \\
DHVC & 112.46 & \textbf{433.81} & \textbf{4.27} & \textbf{0.25} & \textbf{0.21} \\
\bottomrule
\end{tabular}
\caption{{\bf Complexity comparison} among DCVC, VCT, and DHVC (Ours). ``Params" represents the size of model parameters. ``kMACs/pix" denotes the average number of multiply-add operations per pixel. ``PeakMem'' is the peak memory consumed in the inference process. ``ET'' and ``DT'' are the encoding and decoding times. We perform evaluations on a single RTX 3090-24G GPU.}
\label{tbl:complexity}
\end{scriptsize}
\vspace{-0.2cm}
\end{table}

{Our DHVC shows a clear reduction in kMACs/pix and peak memory occupation, owing to the use of simple probabilistic prediction modules instead of complicated Transformer-based prediction network in VCT or complicated motion and residual coding modules in DCVC.} % The increased kMACs/pix of VCT are because of a complex Transformer used for temporal dependency characterization. }
 
{For encoding and decoding time, as the DCVC applies the pixel-wise spatial autoregressive model for entropy coding, it takes about 17.86 and 40.64 seconds, which is unacceptable for practical codecs. VCT, instead, uses a simplified 4$\times$4 block-level spatial autoregressive model, offering faster encoding (decoding) to DCVC. Our method completely removes the use of a spatial autoregressive model through the proposed hierarchical processing pipeline, which further reduces the encoding and decoding time to respective 0.25 and 0.21 seconds, i.e., 55$\times$/221$\times$ (6$\times$/7$\times$) faster encoding/decoding than the DCVC and VCT respectively.}

\begin{figure*}[t]
\centering
\subfloat[]{\includegraphics[width=0.25\linewidth]{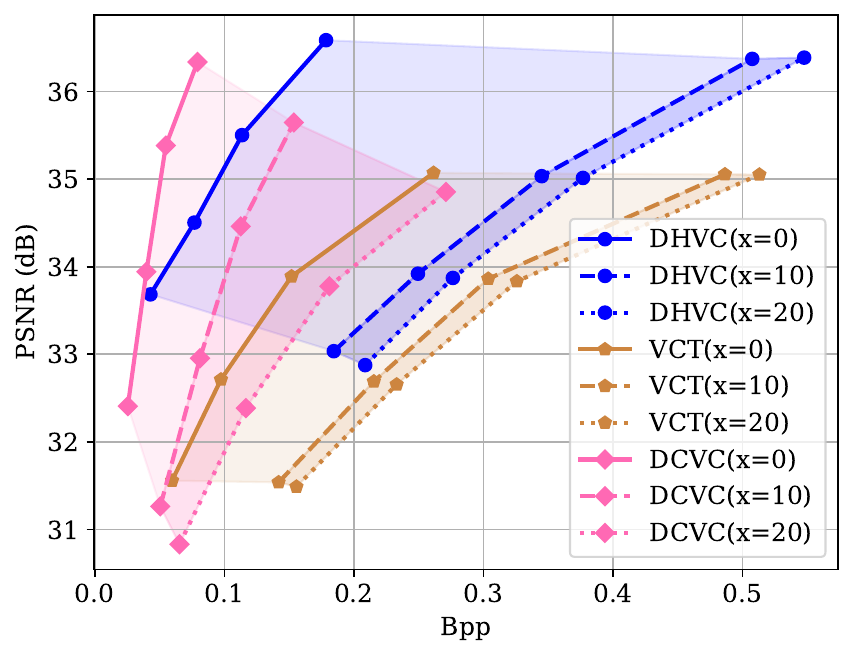}}
\hspace{0.5cm}
\subfloat[]{\includegraphics[width=0.255\linewidth]{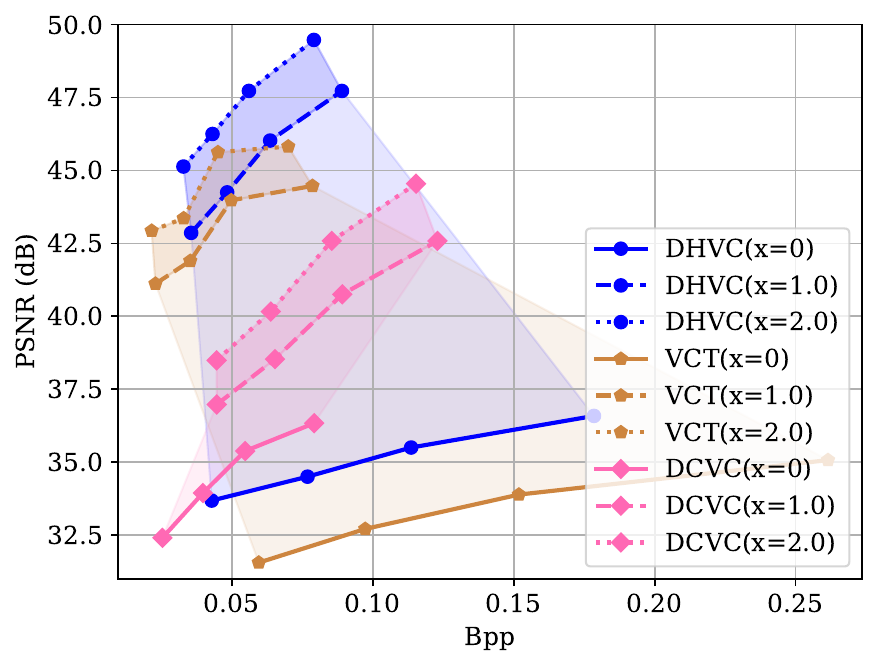}}
\hspace{0.5cm}
\subfloat[]{\includegraphics[width=0.25\linewidth]{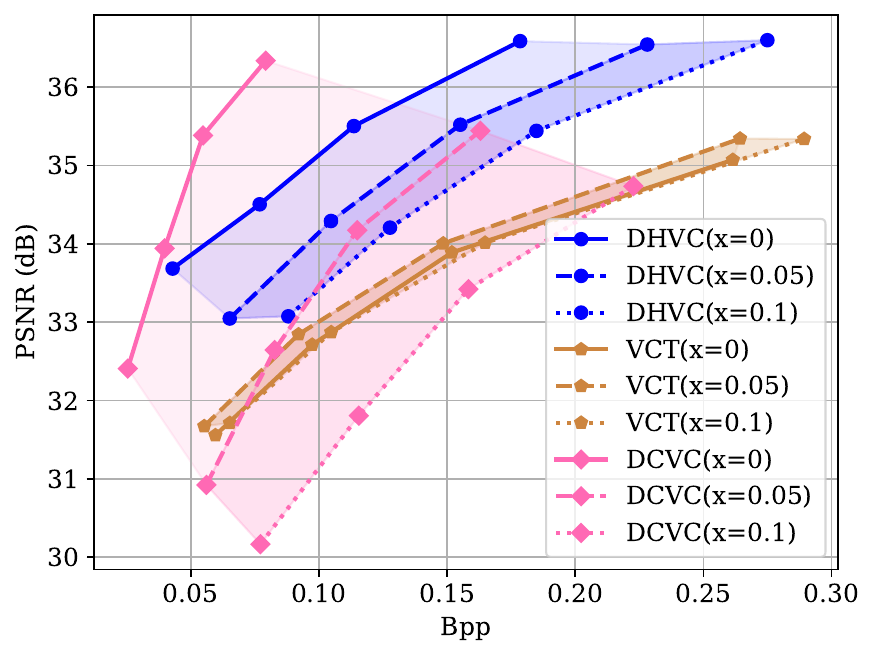}}
\caption{{\bf Impact of temporal pattern} on compression using synthetic data: (a) pixel shifting with values $x=0,10,20$, (b) Gaussian blurring with sigma $x \cdot t$ at frame order $t$, and (c) fading by linear transition between two unrelated scenes using alpha blending. All evaluations start from $x=0$ (i.e., videos consist of still images) as denoted by solid lines in the figures.}
\label{fig:rd_synthetic_data}
\vspace{-0.2cm}
\end{figure*}

\subsection{Deep Dive} \label{sec:ablation_studies}
We perform ablation studies to understand the capacity of our proposed DHVC better. 

\begin{figure*}[htbp]
\centering
\begin{minipage}[t]{0.235\textwidth}
\vspace{0.3cm}
\centering
\includegraphics[width=0.99\linewidth]{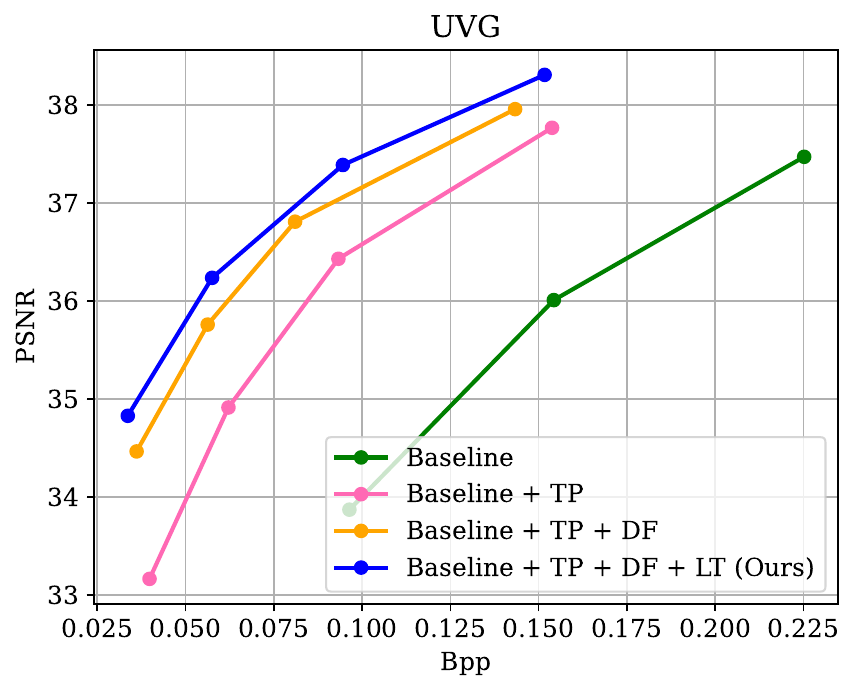}
\caption{{\bf Performance contribution} of modular components.}
\label{fig:ablation}
\end{minipage}
\hspace{8mm}
\begin{minipage}[t]{0.7\textwidth}
\vspace{0pt}
\centering
\includegraphics[width=1\textwidth]{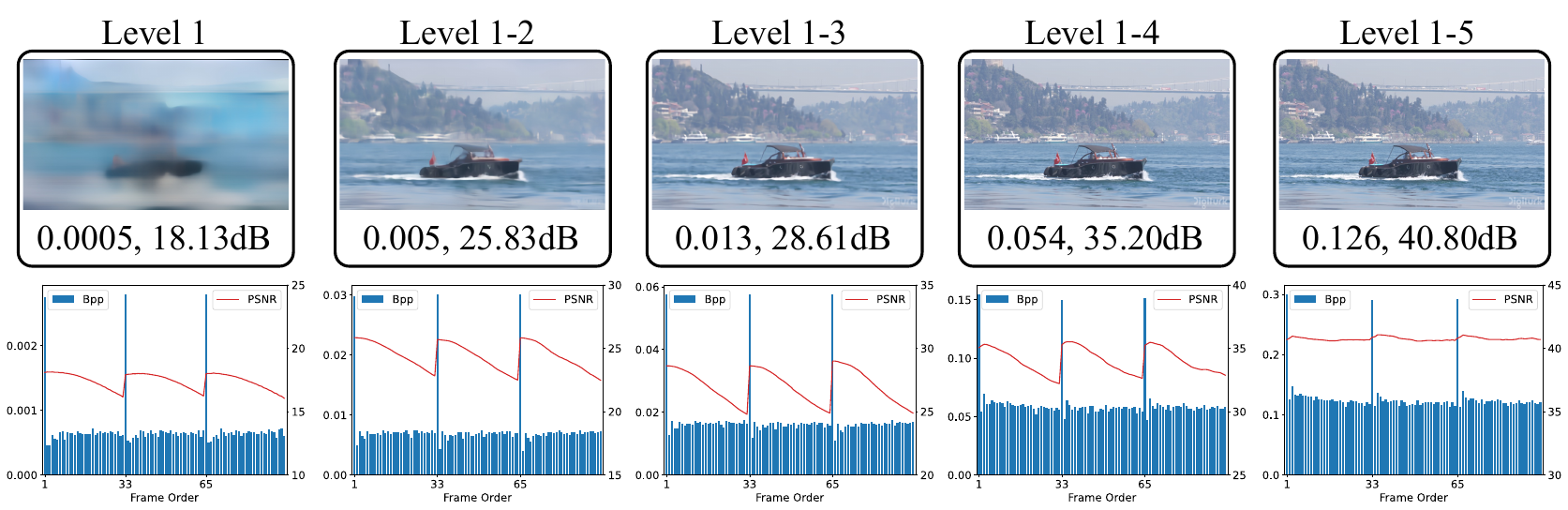}
\caption{{\bf Progressive Decoding.} Five scale levels are visualized. {\it Zoom in} for details. Vertical lines in bottom subplots indicate intra frames for GOP separation.}
\label{fig:progressive}
\end{minipage}
\vspace{-0.2cm}
\end{figure*}

\textbf{Modular Contribution:} 
We further examine the contribution of each module in our proposed DHVC in Fig.~\ref{fig:ablation}. ``Baseline'' denotes the model disabling both the temporal prediction and in-loop decoding fusion in latent blocks, with only the spatial prior from previous scales for probabilistic modeling. ``Baseline + TP'' indicates the temporal probabilistic prediction is integrated to reduce the temporal redundancy. Apparently, the performance with the support of temporal information improves significantly upon the base model. Furthermore, with the help of in-loop decoding fusion module, dubbed by ``Baseline + TP +DF'' in the figure, an averaged 1 dB PSNR increase is obtained. It is sufficient to justify the advantage of compensating for fusion based on prediction results on the decoding side. In addition, the long-term finetuning with five frames, represented by ``Baseline + TP +DF + LT'', brings another R-D improvement to constitute the complete performance of our work. This suggests that the rate-distortion relationship between frames can be effectively balanced by joint training with multiple frames.

% The embedded decoder is then proposed, dubbed by ``Baseline + PP +ED'', to further take advantage of the predictive features. There is averaged more than 1 dB PSNR increase by using such decoder side refinement. Additional long-term finetuning with patches of five frames, termed as ``Baseline + PP +ED + LT'', brings another performance improvement to construct the final capability of our work. Similar as the hybrid motion \& residual coding methods, the rate-distortion relationship of multiple frames can be balanced to achieve the best trade-off in long-term series.

% "w/o PRE" denotes the model without prediction module, which is degradated to the image coding. The compression performance drops dramatically, illustrating the importance of temporal prediction for probabilistic distribution modeling. "w/o DEC" indicates the model without decoded refinement. There is more than 1dB performance gap from our default model, which shows the efficiency of the decoded side refinement. "w/o MFT" represents the model without training using 5 frames window. It seems that the training batch with more frames can further improve the temporal prediction performance.

\textbf{Adaptation Capacity to Temporal Patterns} is critical for the model's generalization when encoding different video contents.
%As a model which is reused for both intra and inter coding, it is interesting to have an experiment to see if the method can be adaptive to the temporal patterns when the proceeding features provide ``noisy" temporal priors. 
We exactly follow \cite{mentzer2022vct} to generate videos using three different temporal patterns, including the pixel shifting, blurring, and fading effects. 
%Specifically, 100 still pictures in the CLIC dataset~\cite{CLIC2020} are selected to make up 100 videos by repeating the picture twelve times. For every temporal pattern, we use the evaluated results using the still videos as the initial performance. We choose the motion shift $x$ of 10 and 20. To simulate the blur pattern, we apply Gaussian blurring with sigma $x \cdot t$ at time step $t$. The fade data is generated by linearly transition between two unrelated images using alpha blending (as in a scene cut). 
The R-D curves are plotted in Fig~\ref{fig:rd_synthetic_data}. Our method outperforms the VCT with regard to all synthetic datasets, which demonstrates the powerful modeling capability of the hierarchical probabilistic predictive mechanism. No matter which temporal pattern or how fast the scene change is, our method is consistently applicable. However, both the VCT and our DHVC behave worse than the DCVC when it comes to the videos with pixel-shifted. This is mainly due to the fact that the DCVC utilizes a motion alignment module by encoding the motion data. For such regular object displacement, motion estimation can achieve high prediction accuracy, which has obvious advantages over our latent space probabilistic prediction. Considering how to add hierarchical motion alignment to our approach is a topic worth future exploring.

% \textcolor{yellow}{The R-D curves are plotted in Fig~\ref{fig:rd_synthetic_data}. With regard to the shift data, the DCVC with motion compensation modules proves to be better than the latent probabilistic methods, i.e., the VCT and ours. However, for the synthetic blur dataset, our method presents better RD performance than both DCVC and VCT. For the fade dataset, our method also shows superior generalization than the DCVC and VCT. }

{\bf Progressive Decoding Capability} is enabled in the proposed DHVC, which was seldom supported in existing methods.    Specifically, once obtain the lowest-scale features (lowest resolution) of the current frame, we have the coarsest frame reconstruction after decoding as in the upper left subplot of Fig.~\ref{fig:progressive} (e.g., Level 1). As additional compressed latent features are transmitted to the decoder side, we can clearly observe the improvement of reconstruction results (see visualized subplots with more scale levels and PSNR increases in bottom subplots accordingly).
When we receive partial scales, we notice the PSNR degradation (red curves) in a GOP. This is due to the error propagation since the temporal references can only provide partial priors for decoding. %This is because the scale-wise conditional probability estimation uses lower-scale spatial and same-scale temporal priors.
Instead, once we receive all-scale latent features, the PSNR metric is stable across frames and GOPs (see ``Level 1 - 5''). At the same time, gradual PSNR degradation in a GOP still prevails in existing video codecs using hybrid motion \& residual coding.

%the current frame will present a better reconstruction result (as seen from the vertical in the figure). On the other hand, the hidden features that are incompletely decoded can still be used as a reference for predictive coding of subsequent frames and will not have impact on the compression performance. 

Progressive decoding quickly provides relatively-coarse reconstructions by encoding and transmitting partial features. In this exemplified Fig.~\ref{fig:progressive}, for such a 1080p video, having two scales of compressed latent features can already present a clear preview of the content, by which our model provides a fast and less bitrate-consuming preview in video streaming applications.  This also gives us a broader understanding: we can still decode the content in networked applications with packet loss when we have partial packets. In congested connections, we can proactively drop latent packets corresponding to higher scales. %Wheras, such a functionality cannot be offered by using 
%Our proposed DHVC can support certain practical functionality extensions due to its hierarchical probabilistic prediction characteristics.

%\textbf{Progressive Decoding:}
%With the properties of hierarchical coding, our model can provide progressive decoding. 

%\textbf{Anti Packet Loss:}
%Furthermore, this also shows that our proposed approach can support a certain level of packet loss resistance. For instance, if the last coded feature of the first frame is lost, the remaining frames can still be decoded correctly by using the first four coded features. In conventional coding standards or previous deep learning-based video coding, packet loss is often fatal. A missing bit as small as 1 bit will cause the current frame, and subsequent frames, to not be decoded properly at all. Our approach offers for the first time the possibility to resist packet loss.

\section{Conclusion} \label{sec:conclusion}
This paper proposed a novel hierarchical probabilistic predictive coding framework for learning-based video compression, termed DHVC. The DHVC provides superior compression efficiency to popular and representative learned video codecs across a great variety of video samples. More importantly, DHVC offers the fastest encoding and decoding with the least running memory, which not only reveals the best balance between coding performance and complexity efficacy but also offers encouraging potential for application of learned video codecs. Our future work will focus on exploring efficient prior representations or optimization mechanisms to further improve the compression efficiency.

% However, as reported in excessive experiments, the inferior performance of the probabilistic predictive coding to traditional codecs (e.g., HEVC or VVC) especially on low-resolution videos is still present, continuously urging the community to explore efficient representation or optimization mechanisms to tackle it, which will be the focus of our future work.

\section*{Acknowledgements}
This work is partially supported by the Natural Science Foundation of China (No.U20A20184). 

\bibliography{aaai24}

\begin{thebibliography}{39}
\providecommand{\natexlab}[1]{#1}

\bibitem[{Agustsson and Theis(2020)}]{agustsson2020universally}
Agustsson, E.; and Theis, L. 2020.
\newblock Universally Quantized Neural Compression.
\newblock \emph{Advances in Neural Information Processing Systems}, 33:
  12367--12376.

\bibitem[{Ball{\'e}, Laparra, and Simoncelli(2016)}]{balle2016end}
Ball{\'e}, J.; Laparra, V.; and Simoncelli, E.~P. 2016.
\newblock End-to-end optimized image compression.
\newblock \emph{arXiv preprint arXiv:1611.01704}.

\bibitem[{Ball{\'e} et~al.(2018)Ball{\'e}, Minnen, Singh, Hwang, and
  Johnston}]{balle2018variational}
Ball{\'e}, J.; Minnen, D.; Singh, S.; Hwang, S.~J.; and Johnston, N. 2018.
\newblock Variational image compression with a scale hyperprior.
\newblock \emph{arXiv preprint arXiv:1802.01436}.

\bibitem[{Bossen et~al.(2013)}]{bossen2013common}
Bossen, F.; et~al. 2013.
\newblock Common test conditions and software reference configurations.
\newblock \emph{JCTVC-L1100}, 12(7): 1.

\bibitem[{Bross et~al.(2021)Bross, Wang, Ye, Liu, Chen, Sullivan, and
  Ohm}]{bross2021overview}
Bross, B.; Wang, Y.-K.; Ye, Y.; Liu, S.; Chen, J.; Sullivan, G.~J.; and Ohm,
  J.-R. 2021.
\newblock Overview of the versatile video coding (VVC) standard and its
  applications.
\newblock \emph{IEEE Transactions on Circuits and Systems for Video
  Technology}, 31(10): 3736--3764.

\bibitem[{Chen et~al.(2022)Chen, Gu, Lu, and Xu}]{chen2022exploiting}
Chen, Z.; Gu, S.; Lu, G.; and Xu, D. 2022.
\newblock Exploiting intra-slice and inter-slice redundancy for learning-based
  lossless volumetric image compression.
\newblock \emph{IEEE Transactions on Image Processing}, 31: 1697--1707.

\bibitem[{Child(2020)}]{child2020very}
Child, R. 2020.
\newblock Very deep vaes generalize autoregressive models and can outperform
  them on images.
\newblock \emph{arXiv preprint arXiv:2011.10650}.

\bibitem[{Duan et~al.(2023)Duan, Lu, Ma, and Zhu}]{duan2023lossy}
Duan, Z.; Lu, M.; Ma, Z.; and Zhu, F. 2023.
\newblock Lossy Image Compression with Quantized Hierarchical VAEs.
\newblock In \emph{Proceedings of the IEEE/CVF Winter Conference on
  Applications of Computer Vision}, 198--207.

\bibitem[{He et~al.(2022)He, Yang, Peng, Ma, Qin, and Wang}]{he2022elic}
He, D.; Yang, Z.; Peng, W.; Ma, R.; Qin, H.; and Wang, Y. 2022.
\newblock Elic: Efficient learned image compression with unevenly grouped
  space-channel contextual adaptive coding.
\newblock In \emph{Proceedings of the IEEE/CVF Conference on Computer Vision
  and Pattern Recognition}, 5718--5727.

\bibitem[{Hu, Lu, and Xu(2021)}]{hu2021fvc}
Hu, Z.; Lu, G.; and Xu, D. 2021.
\newblock FVC: A new framework towards deep video compression in feature space.
\newblock In \emph{Proceedings of the IEEE/CVF Conference on Computer Vision
  and Pattern Recognition}, 1502--1511.

\bibitem[{Kingma and Ba(2014)}]{kingma2014adam}
Kingma, D.~P.; and Ba, J. 2014.
\newblock Adam: A method for stochastic optimization.
\newblock \emph{arXiv preprint arXiv:1412.6980}.

\bibitem[{Kingma et~al.(2016)Kingma, Salimans, Jozefowicz, Chen, Sutskever, and
  Welling}]{kingma2016improved}
Kingma, D.~P.; Salimans, T.; Jozefowicz, R.; Chen, X.; Sutskever, I.; and
  Welling, M. 2016.
\newblock Improved variational inference with inverse autoregressive flow.
\newblock \emph{Advances in neural information processing systems}, 29.

\bibitem[{Li, Li, and Lu(2021)}]{li2021deep}
Li, J.; Li, B.; and Lu, Y. 2021.
\newblock Deep contextual video compression.
\newblock \emph{Advances in Neural Information Processing Systems}, 34:
  18114--18125.

\bibitem[{Li, Li, and Lu(2022)}]{li2022hybrid}
Li, J.; Li, B.; and Lu, Y. 2022.
\newblock Hybrid spatial-temporal entropy modelling for neural video
  compression.
\newblock In \emph{Proceedings of the 30th ACM International Conference on
  Multimedia}, 1503--1511.

\bibitem[{Lin et~al.(2020)Lin, Liu, Li, and Wu}]{lin2020m}
Lin, J.; Liu, D.; Li, H.; and Wu, F. 2020.
\newblock M-LVC: Multiple frames prediction for learned video compression.
\newblock In \emph{Proceedings of the IEEE/CVF Conference on Computer Vision
  and Pattern Recognition}, 3546--3554.

\bibitem[{Liu et~al.(2020{\natexlab{a}})Liu, Lu, Ma, Wang, Xie, Cao, and
  Wang}]{liu2020neural}
Liu, H.; Lu, M.; Ma, Z.; Wang, F.; Xie, Z.; Cao, X.; and Wang, Y.
  2020{\natexlab{a}}.
\newblock Neural video coding using multiscale motion compensation and
  spatiotemporal context model.
\newblock \emph{IEEE Transactions on Circuits and Systems for Video
  Technology}, 31(8): 3182--3196.

\bibitem[{Liu et~al.(2020{\natexlab{b}})Liu, Wang, Ma, Shah, Hu, Dhawan, and
  Urtasun}]{liu2020conditional}
Liu, J.; Wang, S.; Ma, W.-C.; Shah, M.; Hu, R.; Dhawan, P.; and Urtasun, R.
  2020{\natexlab{b}}.
\newblock Conditional entropy coding for efficient video compression.
\newblock In \emph{European Conference on Computer Vision}, 453--468. Springer.

\bibitem[{Liu et~al.(2022)Liu, Mao, Wu, Feichtenhofer, Darrell, and
  Xie}]{liu2022convnet}
Liu, Z.; Mao, H.; Wu, C.-Y.; Feichtenhofer, C.; Darrell, T.; and Xie, S. 2022.
\newblock A convnet for the 2020s.
\newblock In \emph{Proceedings of the IEEE/CVF conference on computer vision
  and pattern recognition}, 11976--11986.

\bibitem[{Lu et~al.(2019)Lu, Ouyang, Xu, Zhang, Cai, and Gao}]{lu2019dvc}
Lu, G.; Ouyang, W.; Xu, D.; Zhang, X.; Cai, C.; and Gao, Z. 2019.
\newblock Dvc: An end-to-end deep video compression framework.
\newblock In \emph{Proceedings of the IEEE/CVF Conference on Computer Vision
  and Pattern Recognition}, 11006--11015.

\bibitem[{Lu et~al.(2020)Lu, Zhang, Ouyang, Chen, Gao, and Xu}]{lu2020end}
Lu, G.; Zhang, X.; Ouyang, W.; Chen, L.; Gao, Z.; and Xu, D. 2020.
\newblock An end-to-end learning framework for video compression.
\newblock \emph{IEEE transactions on pattern analysis and machine
  intelligence}, 43(10): 3292--3308.

\bibitem[{Lu et~al.(2022)Lu, Guo, Shi, Cao, and Ma}]{9810760}
Lu, M.; Guo, P.; Shi, H.; Cao, C.; and Ma, Z. 2022.
\newblock Transformer-based Image Compression.
\newblock In \emph{2022 Data Compression Conference (DCC)}, 469--469.

\bibitem[{Marcellin et~al.(2000)Marcellin, Gormish, Bilgin, and
  Boliek}]{marcellin2000overview}
Marcellin, M.~W.; Gormish, M.~J.; Bilgin, A.; and Boliek, M.~P. 2000.
\newblock An overview of JPEG-2000.
\newblock In \emph{Proceedings DCC 2000. Data Compression Conference},
  523--541. IEEE.

\bibitem[{Mentzer et~al.(2022)Mentzer, Toderici, Minnen, Hwang, Caelles, Lucic,
  and Agustsson}]{mentzer2022vct}
Mentzer, F.; Toderici, G.; Minnen, D.; Hwang, S.-J.; Caelles, S.; Lucic, M.;
  and Agustsson, E. 2022.
\newblock Vct: A video compression transformer.
\newblock \emph{arXiv preprint arXiv:2206.07307}.

\bibitem[{Mercat, Viitanen, and Vanne(2020)}]{mercat2020uvg}
Mercat, A.; Viitanen, M.; and Vanne, J. 2020.
\newblock UVG dataset: 50/120fps 4K sequences for video codec analysis and
  development.
\newblock In \emph{Proceedings of the 11th ACM Multimedia Systems Conference},
  297--302.

\bibitem[{Ryder et~al.(2022)Ryder, Zhang, Kang, and
  Zhang}]{ryder2022split_hierarchical}
Ryder, T.; Zhang, C.; Kang, N.; and Zhang, S. 2022.
\newblock Split Hierarchical Variational Compression.
\newblock \emph{Proceedings of the IEEE/CVF Conference on Computer Vision and
  Pattern Recognition)}, 386--395.

\bibitem[{Sheng et~al.(2022)Sheng, Li, Li, Li, Liu, and Lu}]{9941493}
Sheng, X.; Li, J.; Li, B.; Li, L.; Liu, D.; and Lu, Y. 2022.
\newblock Temporal Context Mining for Learned Video Compression.
\newblock \emph{IEEE Transactions on Multimedia}, 1--12.

\bibitem[{Sullivan et~al.(2012)Sullivan, Ohm, Han, and
  Wiegand}]{sullivan2012overview}
Sullivan, G.~J.; Ohm, J.-R.; Han, W.-J.; and Wiegand, T. 2012.
\newblock Overview of the high efficiency video coding (HEVC) standard.
\newblock \emph{IEEE Transactions on circuits and systems for video
  technology}, 22(12): 1649--1668.

\bibitem[{Theis and Ahmed(2022)}]{theis2022algorithms_comm_sapmles}
Theis, L.; and Ahmed, N.~Y. 2022.
\newblock Algorithms for the Communication of Samples.
\newblock \emph{Proceedings of the International Conference on Machine
  Learning}, 162: 21308--21328.

\bibitem[{Theis et~al.(2017)Theis, Shi, Cunningham, and
  Husz{\'a}r}]{theis2017lossy}
Theis, L.; Shi, W.; Cunningham, A.; and Husz{\'a}r, F. 2017.
\newblock Lossy Image Compression with Compressive Autoencoders.
\newblock \emph{International Conference on Learning Representations}.

\bibitem[{Vahdat and Kautz(2020)}]{vahdat2020nvae}
Vahdat, A.; and Kautz, J. 2020.
\newblock NVAE: A deep hierarchical variational autoencoder.
\newblock \emph{Advances in neural information processing systems}, 33:
  19667--19679.

\bibitem[{Wallace(1991)}]{wallace1991jpeg}
Wallace, G.~K. 1991.
\newblock The JPEG still picture compression standard.
\newblock \emph{Communications of the ACM}, 34(4): 30--44.

\bibitem[{Wang et~al.(2016)Wang, Gan, Hu, Lin, Jin, Song, Wang, Katsavounidis,
  Aaron, and Kuo}]{wang2016mcl}
Wang, H.; Gan, W.; Hu, S.; Lin, J.~Y.; Jin, L.; Song, L.; Wang, P.;
  Katsavounidis, I.; Aaron, A.; and Kuo, C.-C.~J. 2016.
\newblock MCL-JCV: a JND-based H. 264/AVC video quality assessment dataset.
\newblock In \emph{2016 IEEE international conference on image processing
  (ICIP)}, 1509--1513. IEEE.

\bibitem[{Wang, Simoncelli, and Bovik(2003)}]{wang2003multiscale}
Wang, Z.; Simoncelli, E.~P.; and Bovik, A.~C. 2003.
\newblock Multiscale structural similarity for image quality assessment.
\newblock In \emph{The Thrity-Seventh Asilomar Conference on Signals, Systems
  \& Computers, 2003}, volume~2, 1398--1402. Ieee.

\bibitem[{Wiegand et~al.(2003)Wiegand, Sullivan, Bjontegaard, and
  Luthra}]{wiegand2003overview}
Wiegand, T.; Sullivan, G.~J.; Bjontegaard, G.; and Luthra, A. 2003.
\newblock Overview of the H. 264/AVC video coding standard.
\newblock \emph{IEEE Transactions on circuits and systems for video
  technology}, 13(7): 560--576.

\bibitem[{Xue et~al.(2019)Xue, Chen, Wu, Wei, and Freeman}]{xue2019video}
Xue, T.; Chen, B.; Wu, J.; Wei, D.; and Freeman, W.~T. 2019.
\newblock Video enhancement with task-oriented flow.
\newblock \emph{International Journal of Computer Vision}, 127: 1106--1125.

\bibitem[{Yang et~al.(2020)Yang, Mentzer, Van~Gool, and
  Timofte}]{yang2020learning}
Yang, R.; Mentzer, F.; Van~Gool, L.; and Timofte, R. 2020.
\newblock Learning for video compression with recurrent auto-encoder and
  recurrent probability model.
\newblock \emph{IEEE Journal of Selected Topics in Signal Processing}, 15(2):
  388--401.

\bibitem[{Yang, Bamler, and Mandt(2020{\natexlab{a}})}]{yang2020improving}
Yang, Y.; Bamler, R.; and Mandt, S. 2020{\natexlab{a}}.
\newblock Improving inference for neural image compression.
\newblock \emph{Advances in Neural Information Processing Systems}, 33:
  573--584.

\bibitem[{Yang, Bamler, and Mandt(2020{\natexlab{b}})}]{yang2020quantization}
Yang, Y.; Bamler, R.; and Mandt, S. 2020{\natexlab{b}}.
\newblock Variational Bayesian Quantization.
\newblock \emph{Proceedings of the International Conference on Machine
  Learning}, 119: 10670--10680.

\bibitem[{Yang and Mandt(2022)}]{yang2022sandwich}
Yang, Y.; and Mandt, S. 2022.
\newblock Towards Empirical Sandwich Bounds on the Rate-Distortion Function.
\newblock \emph{International Conference on Learning Representations}.

\end{thebibliography}

\clearpage
\appendix
\section{Appendix}
In this supplementary material, we provide additional implementation details and experimental results for more thorough description and validation of our method. First, we depict the detailed network architecture used for our model layer by layer. Then, we illustrate the actual compression and decompression process. Furthermore, we give the command for encoding with x265 and HM 16.26 as the comparison benchmarks for traditional codecs (i.e., H.265) in the main paper. We also have a rate distribution analysis for latent variable each scale. The enlarged R-D curves for both MSE and MS-SSIM optimized models are presented. In the end, we give more visualizations of the progressive decoding to have a more interpretable illustration. Codes are attached with this document.

\subsection{Details of Network Architecture}

\begin{figure*}[htbp]
\centering
\includegraphics[width=\linewidth]{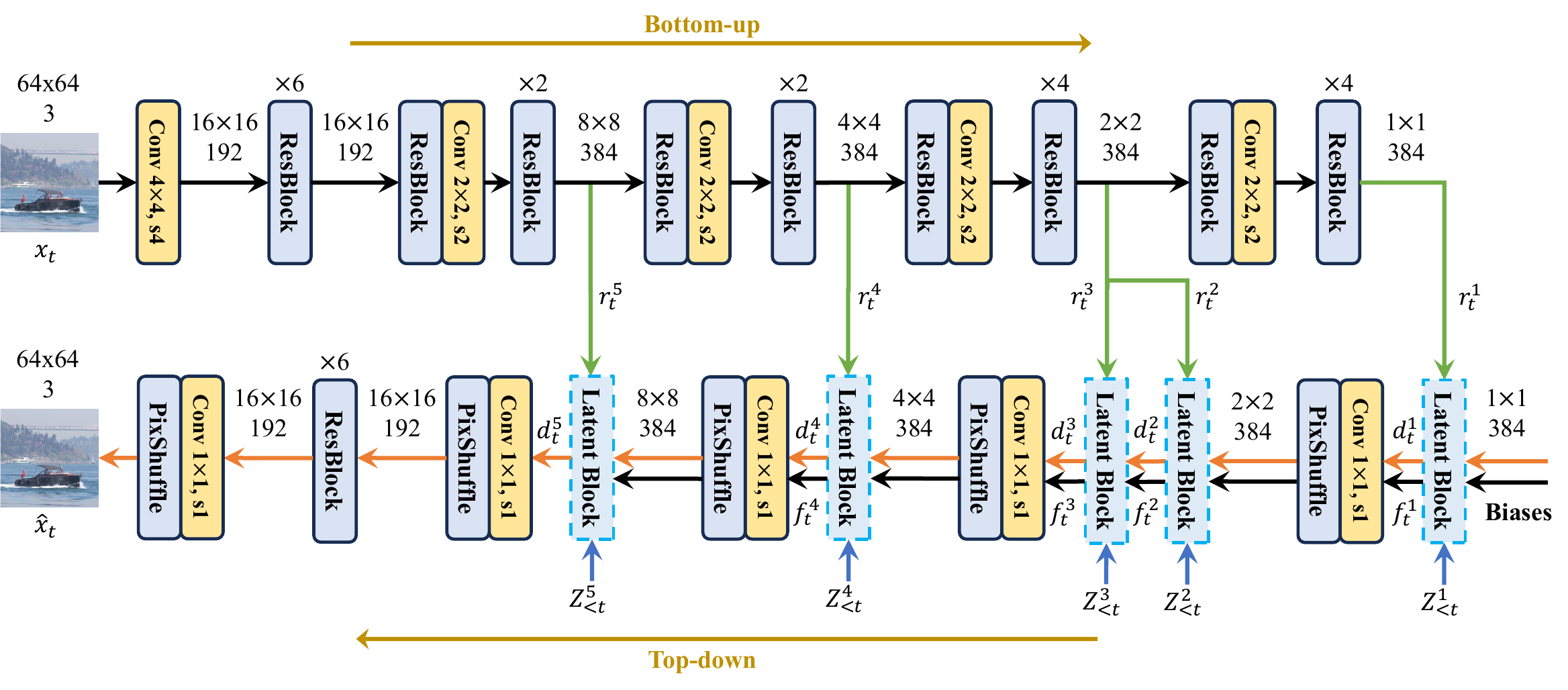}
\caption{{\bf Details of our network architecture.}}
\label{fig:detailed_network}
\end{figure*}

We depict the detailed network architecture in Fig.~\ref{fig:detailed_network}. Take a frame with the resolution of $64 \times 64$ as an example, for which the input channel is 3. For the bottom-up path, the convolutional layer with kernel size at $4 \times 4$ and stride size at 4 (denoted by `s4' in the figure) is used for downsampling to obtain the feature at $16 \times 16$ with 192 channels. Six ResBlocks are used for information aggregation. The following operations are sequentially processed as depicted in the figure. Five features, i.e., $r_t^1, ..., r_t^5$ at different scales are generated to pass into the Latent Blocks at top-down path for encoding. The temporal latent variables ($Z_{<t}^1, ..., Z_{<t}^5$) are employed for probabilistic distribution modeling in Latent Blocks. Two distinct features, i.e., $f_t^l$ and $d_t^l$, are generated by each Latent Block. We use $d_t^5$ to reconstruct the final decoded result $\hat{x}_t$.

\subsection{Compression and Decompression}

As for actual compression and decompression, the framework is basically unchanged, except the latent variables are quantized for entropy coding. At each Latent Block, we use uniform quantization to quantize $\mu_t^l$ from posterior instead of additive uniform noise or straight through quantization at training time. We also discretize the prior to form a discretized Gaussian probability mass function. 

\begin{figure*}[htbp]
\centering
\subfloat[]{\includegraphics[width=0.45\linewidth]{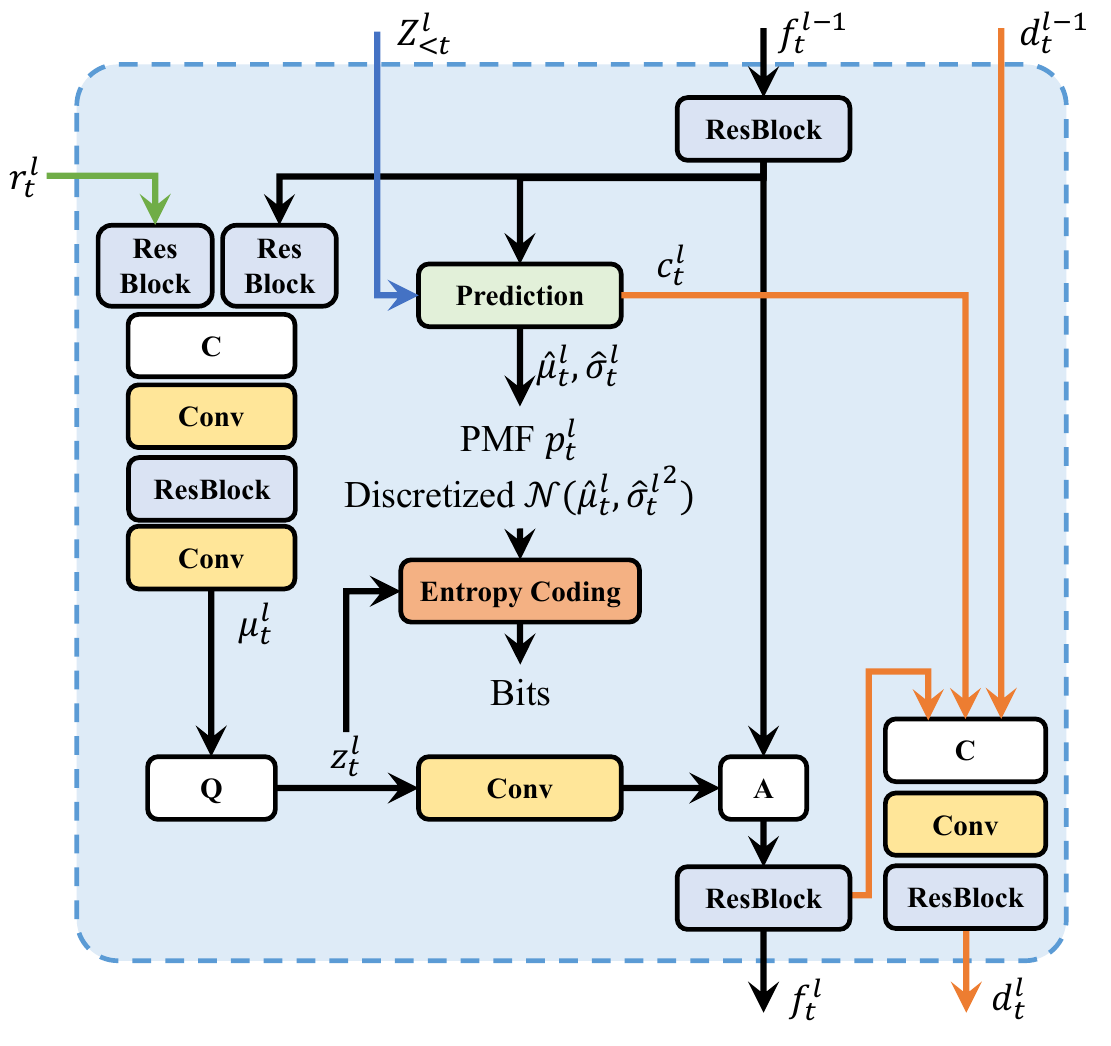}\label{fig:latent_block_compress}}
\hspace{0.5cm}
\subfloat[]{\includegraphics[width=0.355\linewidth]{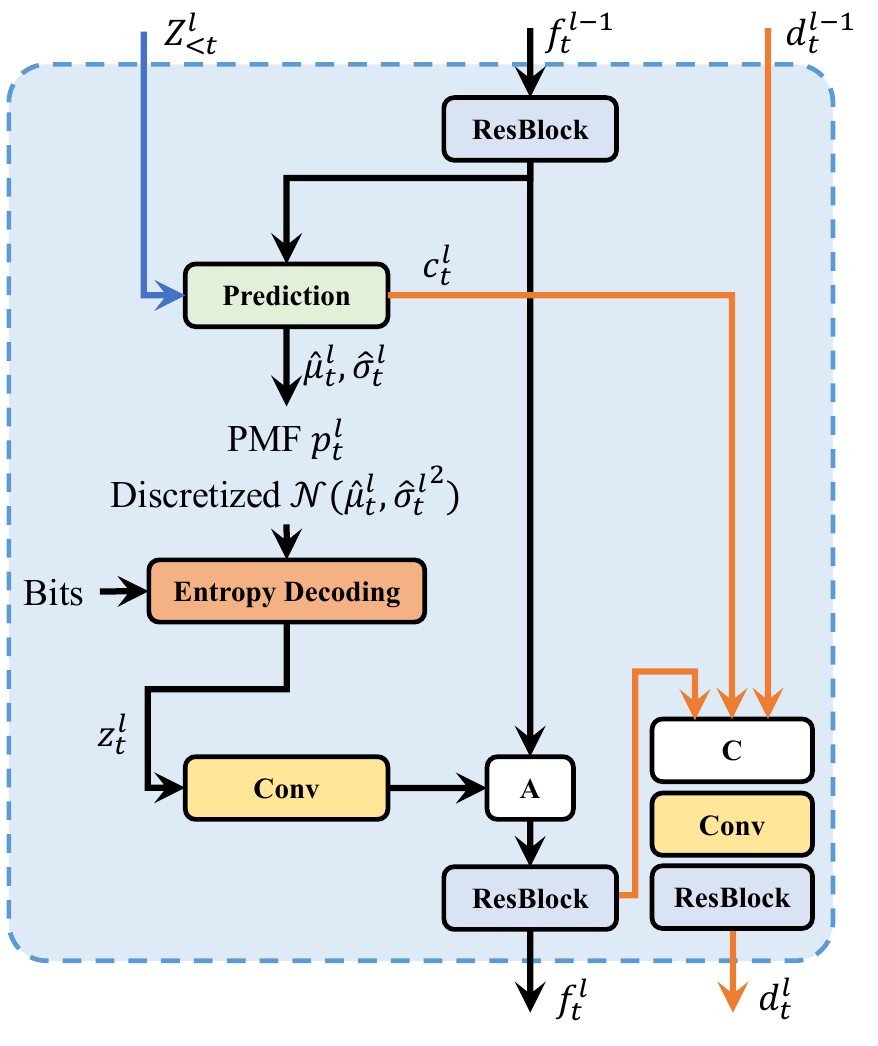}\label{fig:latent_block_decompress}}
\caption{Illustration of the (a) {\bf Compression} and (b) {\bf Decompression} of the Latent Block.}
\end{figure*}

The compression process is detailed in Fig.~\ref{fig:latent_block_compress}. We follow the CompressAI (\url{https://interdigitalinc.github.io/CompressAI/}) by using the residual rounding for each $\mu_t^l$ as below:
\begin{align}
    z_t^l = \hat{\mu}_t^l + \lfloor \mu_t^l - \hat{\mu}_t^l \rceil
\end{align}
where $\lfloor \cdot \rceil$ is the nearest integer rounding function. Then we can encode the quantized $z_t^l$ into bits using the probability mass function (PMF) $p_t^l(\cdot)$. Each of Latent Block produces a separate bitstream, so a compressed frame consists of $L$ bitstreams, corresponding to $L$ latent variables $z_t^1, ..., z_t^L$.

Decompression (Fig.~\ref{fig:latent_block_decompress}) is done in a similar way. Starting from the constant bias, we iteratively compute $p_t^l(\cdot)$ and decode $z_t^l$ from the $l$-th bitstream at each Latent Block. Then the $z_t^l$ is transformed using convolutional layers before added to the prior feature. Once this is done for all $l = 1,2,...,L$, we can obtain the reconstruction $\hat{x}$ using the final upsampling layer in the top-down decoder.

\subsection{Settings for Traditional Codecs}

The command to encode using x265 in our paper is:

\begin{lstlisting}[language=Bash]
ffmpeg
-y
-pix_fmt yuv420p
-s {width}x{height}
-framerate {frame rate}
-i {input file name}
-vframes 96
-c:v libx265
-preset veryslow
-tune zerolatency
-x265-params
``qp={qp}:keyint=32:csv-log-level=1:
csv={csv path}:verbose=1:psnr=1''
{bitstream file name}
\end{lstlisting}

The command to encode using HM 16.26 in our paper is:
\begin{lstlisting}[language=Bash]
TAppEncoder
-c encoder_lowdelay_main_rext.cfg
--InputFile={input file name}
--SourceWidth={width}
--SourceHeight={height}
--InputBitDepth=8
--OutputBitDepth=8
--OutputBitDepthC=8
--InputChromaFormat=444
--FrameRate={frame rate}
--FramesToBeEncoded=96
--IntraPeriod=32
--DecodingRefreshType=2
--QP={qp}
--Level=6.2
--BitstreamFile={bitstream file name}
\end{lstlisting}

\subsection{Rate Distribution}

\begin{figure*}[htbp]
\centering
\subfloat[]{\includegraphics[width=0.8\linewidth]{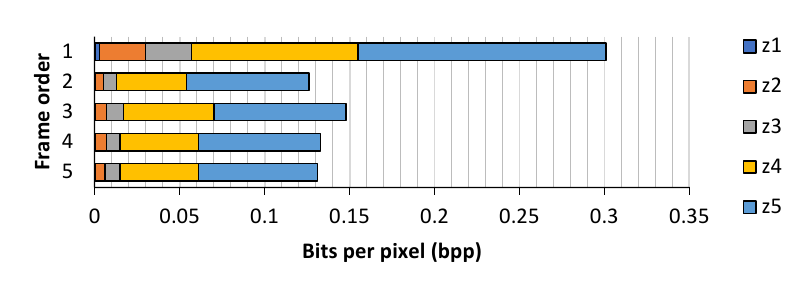} \label{fig:bpp_distribution}} \\
\subfloat[]{\includegraphics[width=0.8\linewidth]{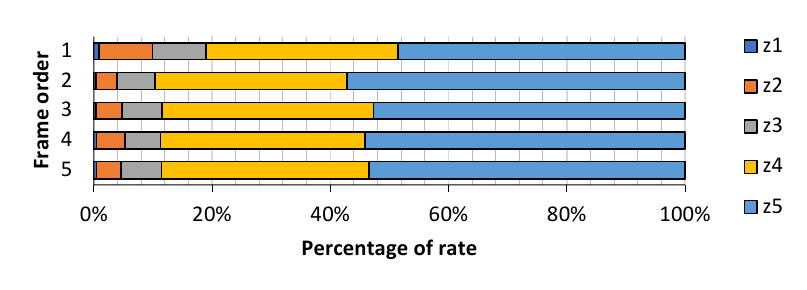} \label{fig:percentage_distribution}}
\caption{{\bf Rate distribution over latent variables} for successive frames within a video sequence. The x-axis is bpp in (a) and percentage in (b). In both figures, rows correspond to different frame orders. In each row, different color represents different latent variables: $z1$ has the smallest spatial dimension ($64\times$ downsampled from the input frame), and $z5$ has the largest spatial dimension ($8\times$ downsampled from the input frame).}
\label{fig:rate_distribution}
\end{figure*}

Recall that the DHVC model produces $L=5$ bitstreams for each frame. Each bitstream corresponds to a different latent variable, and the lengths of the bitstreams vary for different frames. To better understand how the rate distributes over latent variables and how inter-coding impacts such distributions, we visualize them in Fig.~\ref{fig:rate_distribution} using the UVG dataset as an example.

Several interesting facts can be observed from Fig.~\ref{fig:bpp_distribution}. First, we see that the intra frame is with a higher overall rate without the conditional probabilistic modeling. When looking at the rate for each individual latent variable, one can observe that the rates for variables increase progressively and the overall rate is dominated by higher-dimensional latent variables. We also depict in Fig.~\ref{fig:percentage_distribution} the percentage (instead of absolute values) of the rate each latent variable consumes, i.e., the overall rate of each frame. The observation is consistent with Fig.~\ref{fig:bpp_distribution}. The majority of the bit rate is produced by high-dimensional latent variables, regardless of the intra or inter frame.

\subsection{RD Curves}
\begin{figure*}[t]
\centering
\subfloat[]{\includegraphics[width=0.45\linewidth]{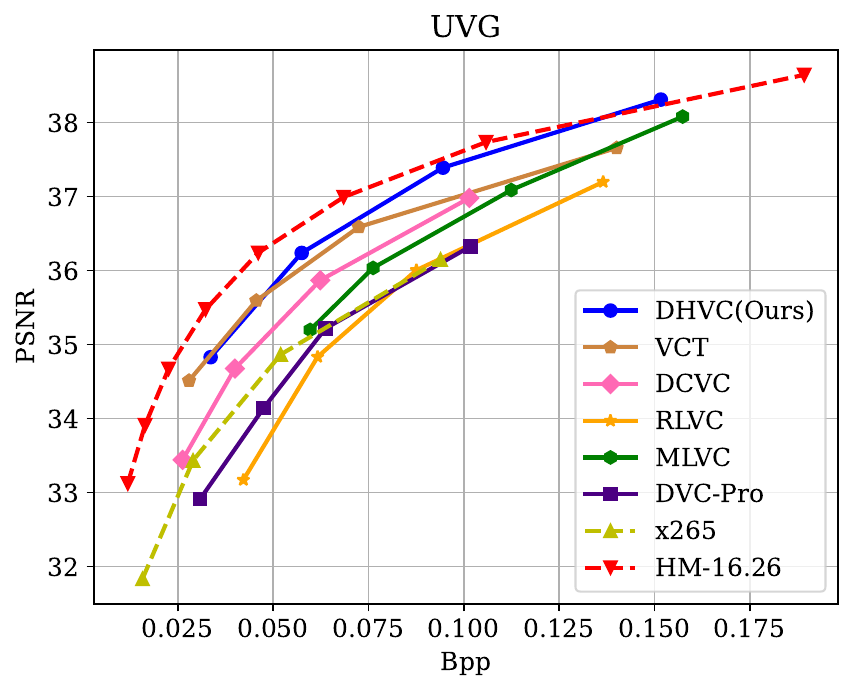}}
\hspace{0.5cm}
\subfloat[]{\includegraphics[width=0.45\linewidth]{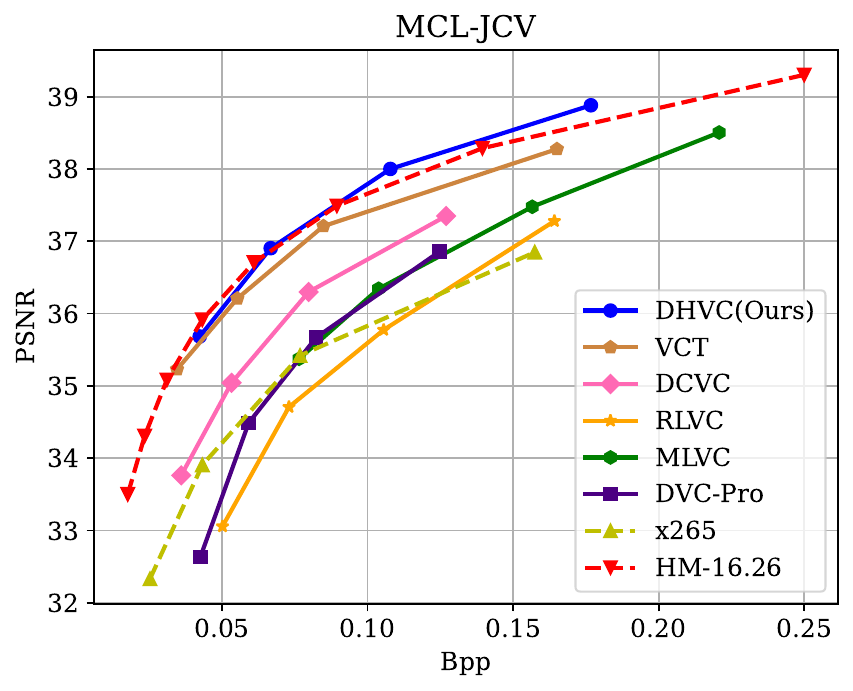}}
\\
\subfloat[]{\includegraphics[width=0.45\linewidth]{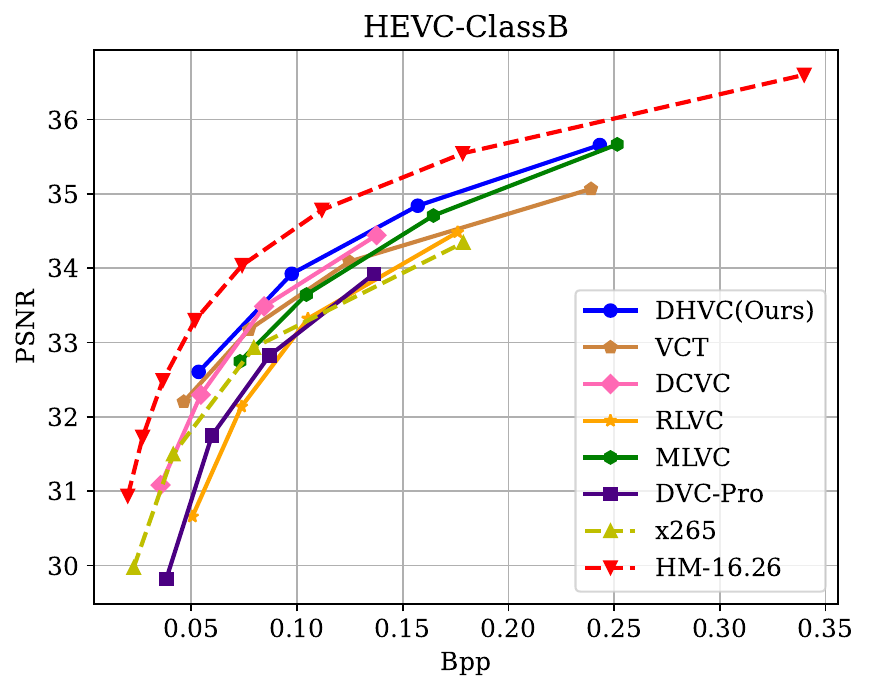}}
\hspace{0.5cm}
\subfloat[]{\includegraphics[width=0.45\linewidth]{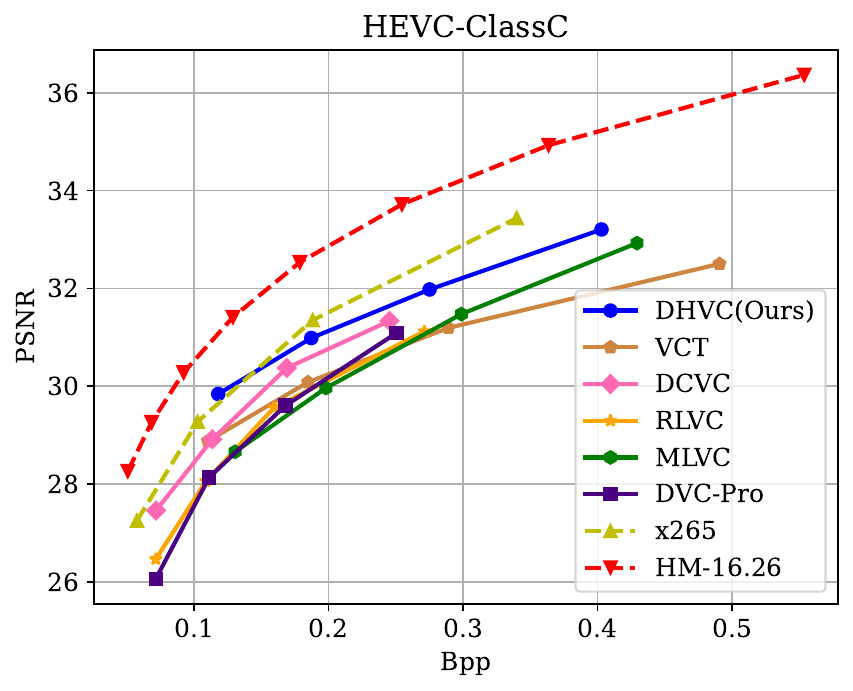}}
\\
\subfloat[]{\includegraphics[width=0.45\linewidth]{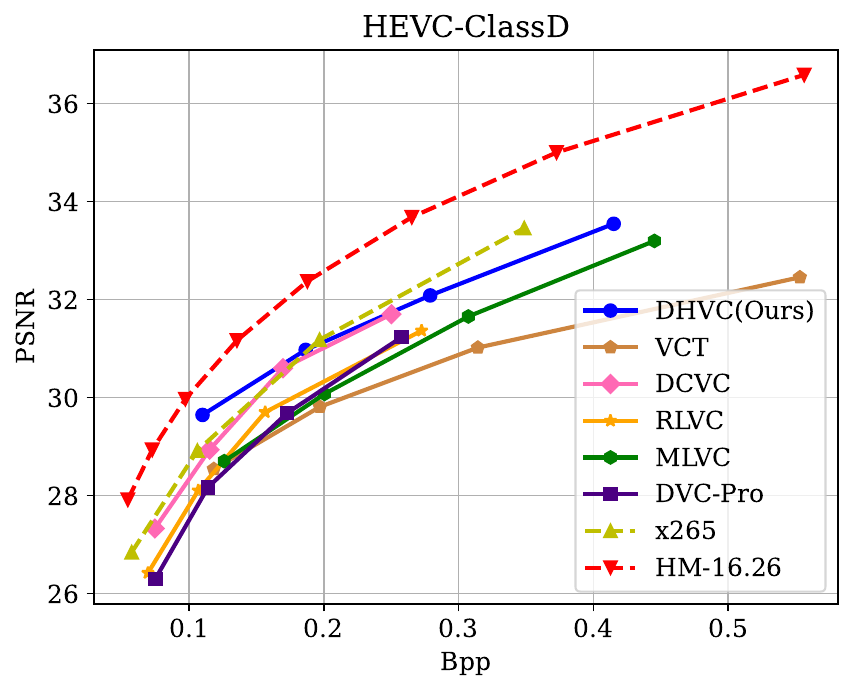}}
\hspace{0.5cm}
\subfloat[]{\includegraphics[width=0.45\linewidth]{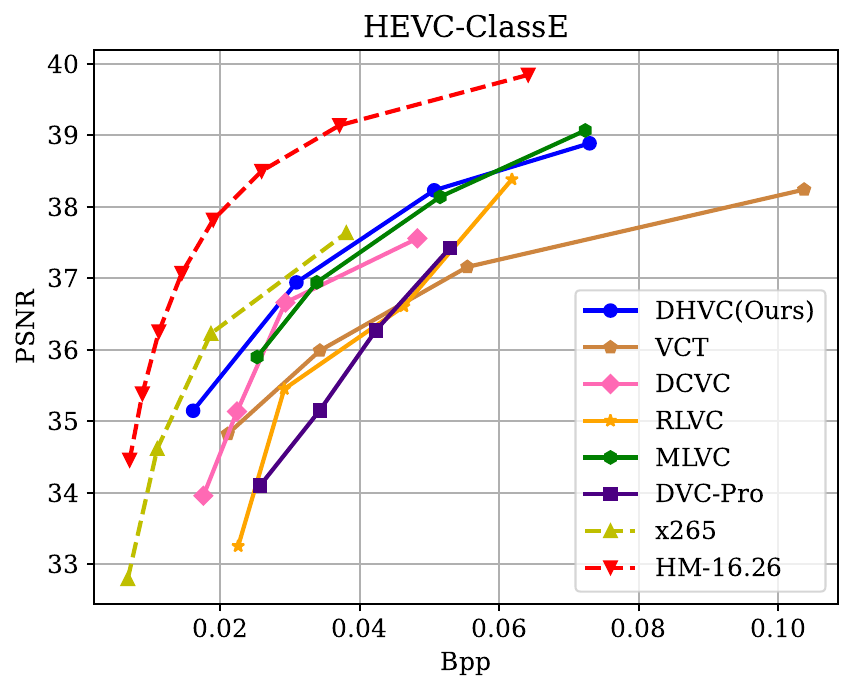}}
\caption{{\bf R-D curves with MSE optimized.}}
\label{fig:large_rd_mse}
\end{figure*}

\begin{figure*}[htbp]
\centering
\subfloat[]{\includegraphics[width=0.45\linewidth]{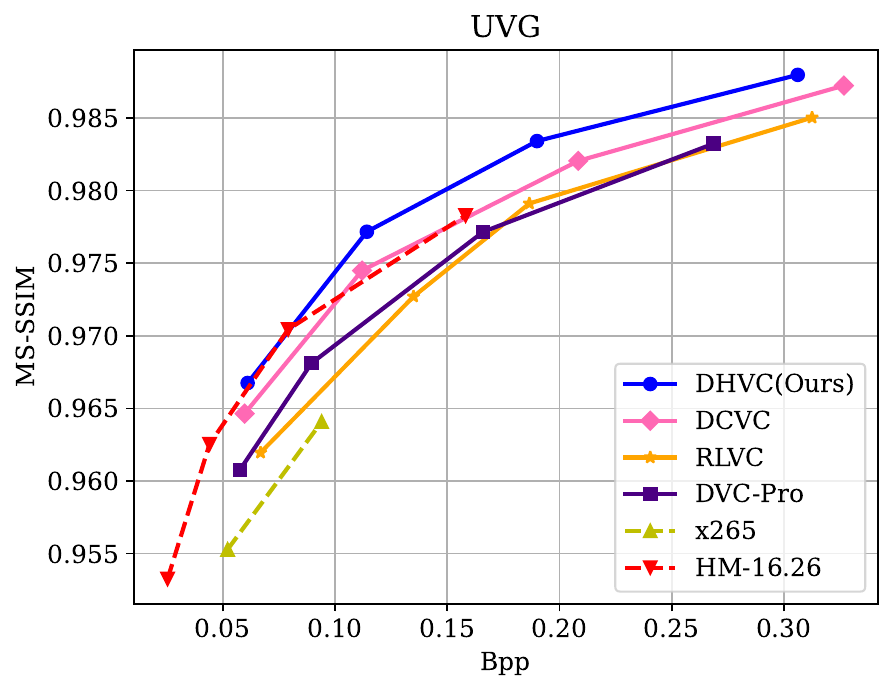}}
\hspace{0.5cm}
\subfloat[]{\includegraphics[width=0.45\linewidth]{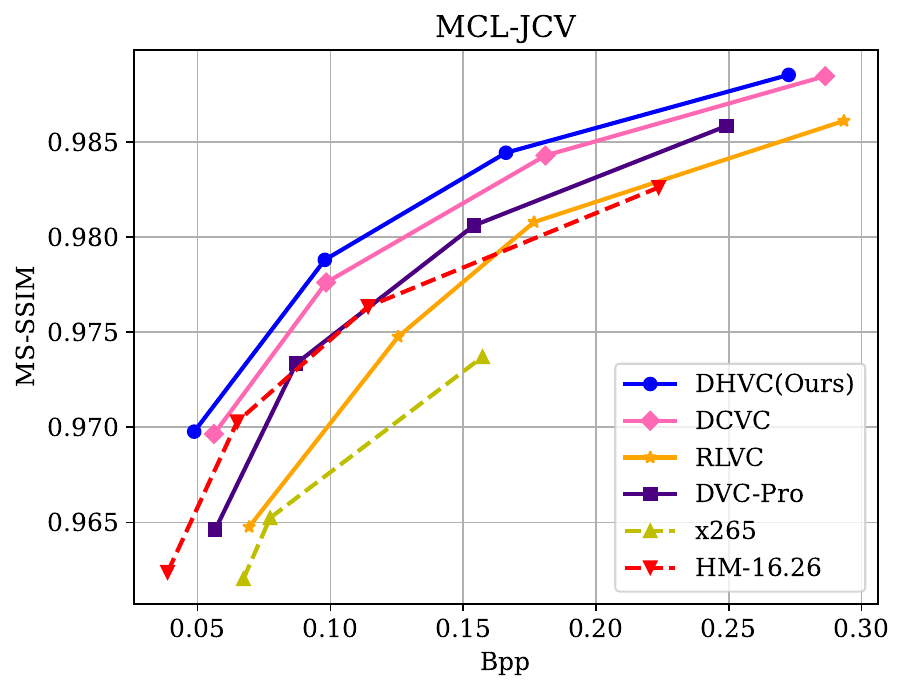}}
\\
\subfloat[]{\includegraphics[width=0.45\linewidth]{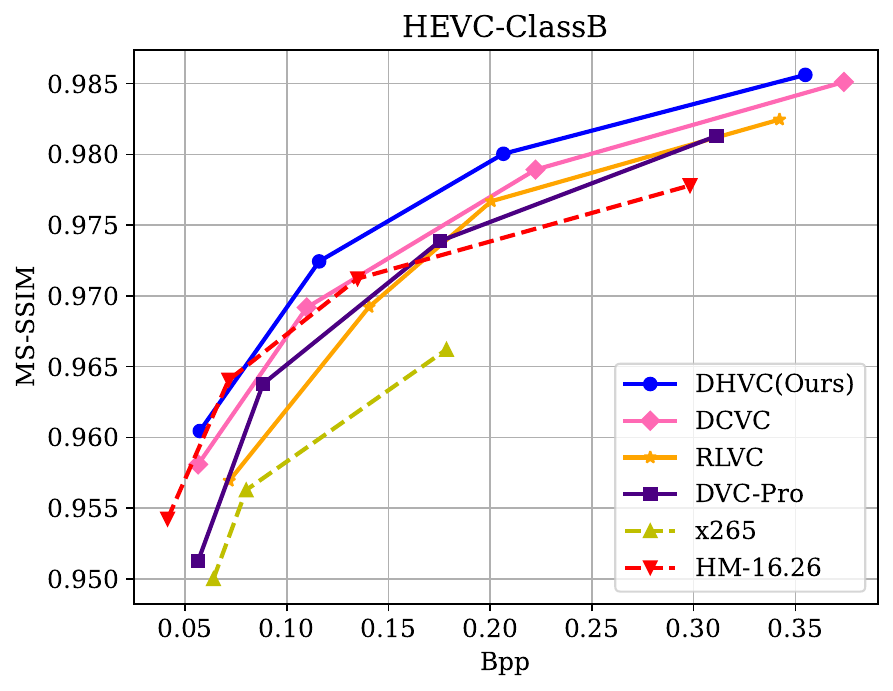}}
\hspace{0.5cm}
\subfloat[]{\includegraphics[width=0.45\linewidth]{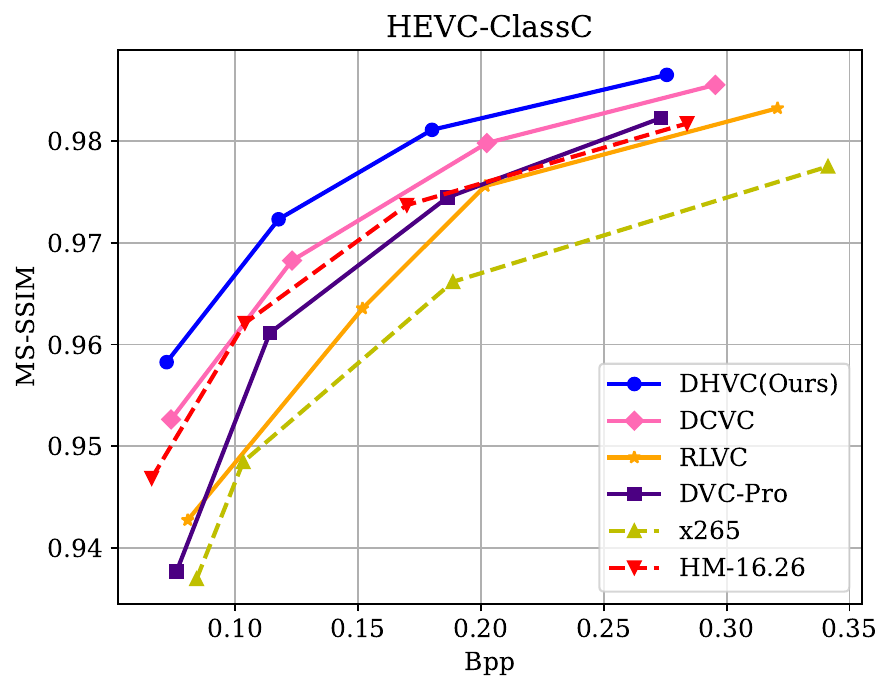}}
\\
\subfloat[]{\includegraphics[width=0.45\linewidth]{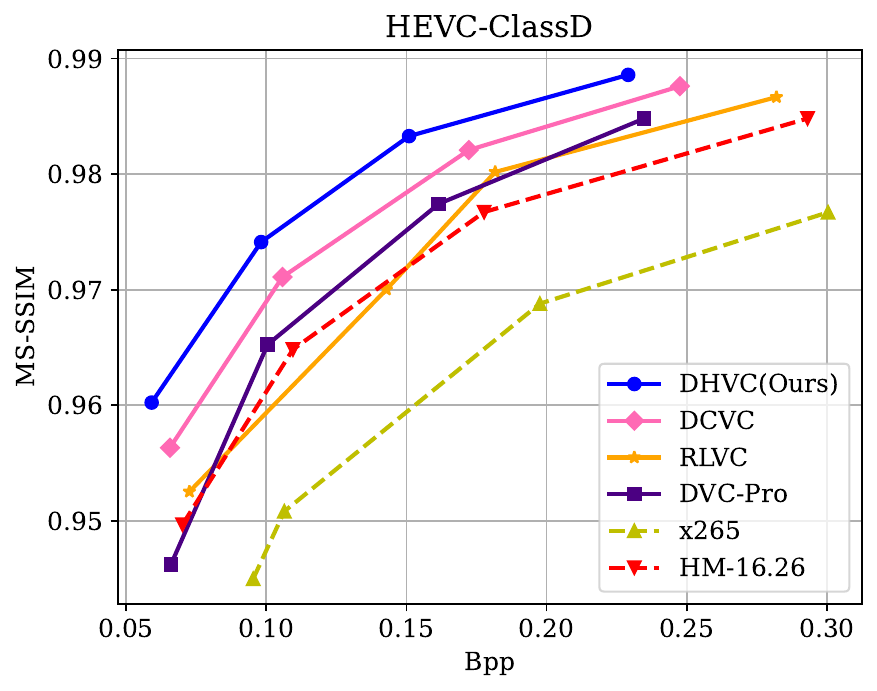}}
\hspace{0.5cm}
\subfloat[]{\includegraphics[width=0.45\linewidth]{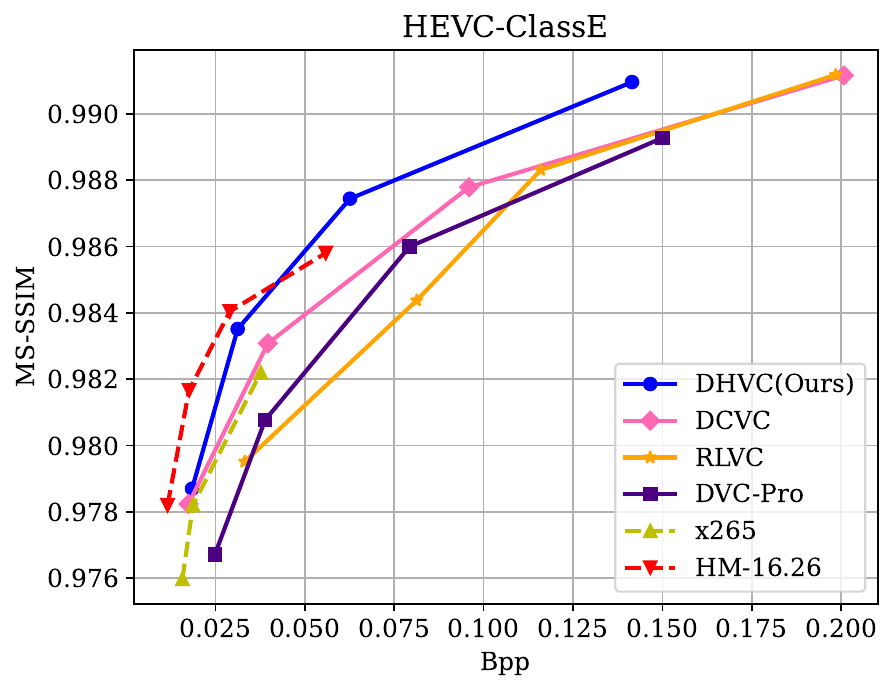}}
\caption{{\bf R-D curves with MS-SSIM optimized.}}
\label{fig:large_rd_msssim}
\end{figure*}

The objective comparison results are shown in Fig.~\ref{fig:large_rd_mse} and Fig.~\ref{fig:large_rd_msssim}. All the implementation settings are kept the same as in the main paper. Noted that the MS-SSIM optimized models are fine-tuned from MSE optimized models for another 100K steps, respectively.

\subsection{More Visualizations}
\begin{figure*}[htbp]
\centering
\includegraphics[width=\linewidth]{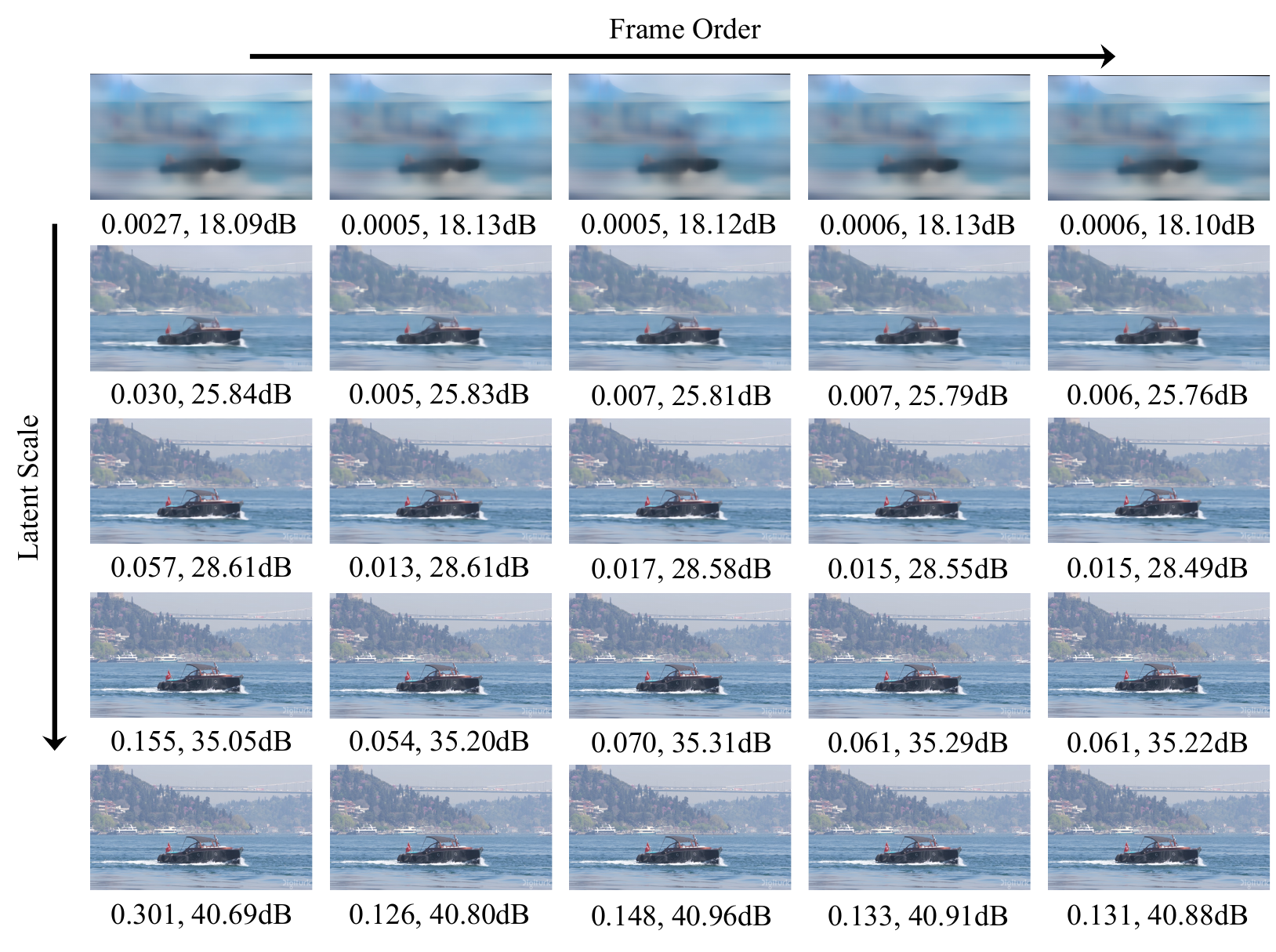}
\caption{{\bf Progressive decoding} of five successive frames using ``Bosphorus'' in UVG dataset for visualization. Please {\it zoom in} for a better view.}
\label{fig:bosphorus_progressive}
\end{figure*}

\begin{figure*}[htbp]
\centering
\includegraphics[width=\linewidth]{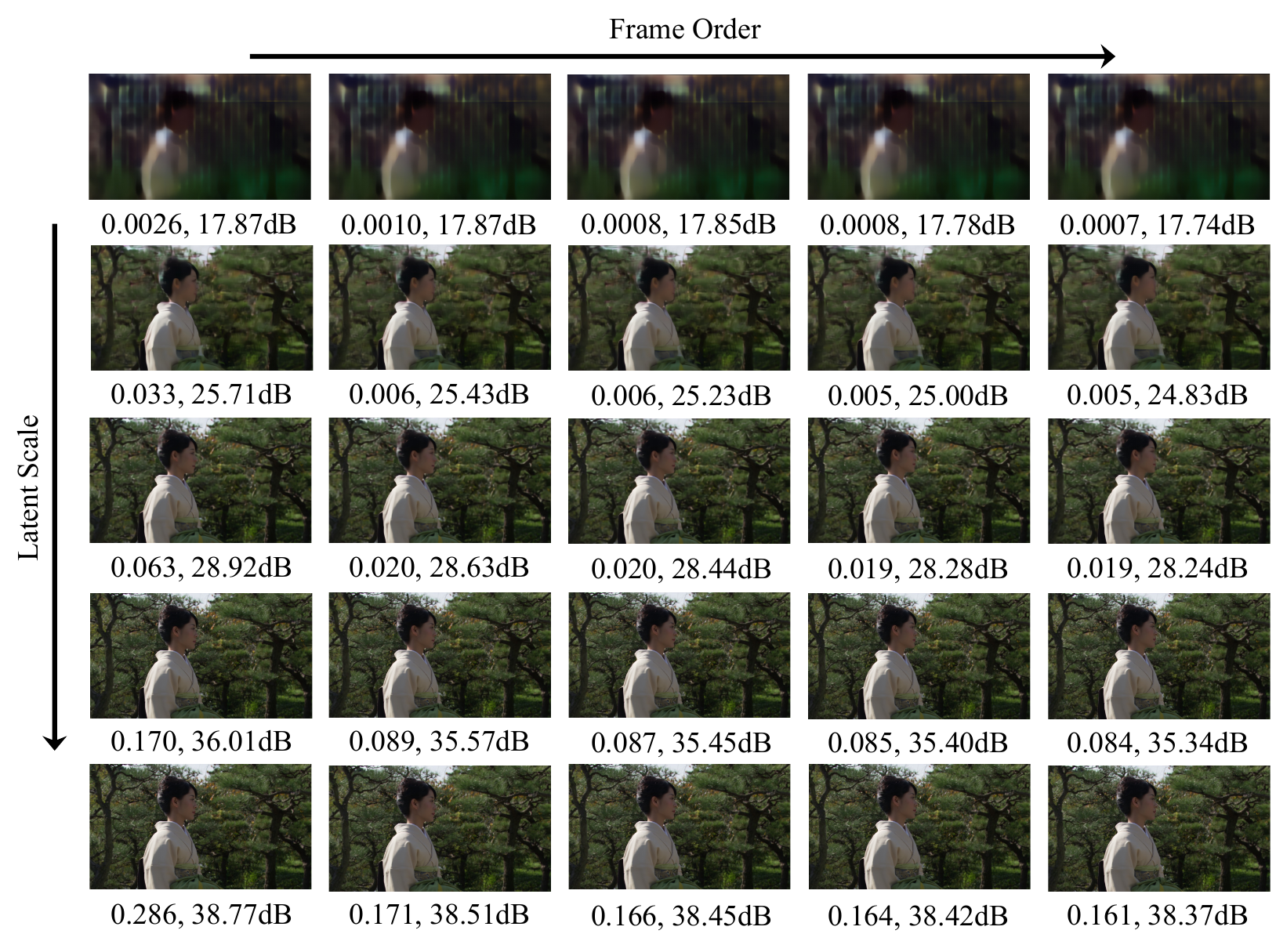}
\caption{{\bf Progressive decoding} of five successive frames using ``videoSRC16'' in MCL-JCV dataset for visualization. Please {\it zoom in} for a better view.}
\label{fig:videoSRC16_progressive}
\end{figure*}

Two video sequences, i.e., ``Bosphorus'' in UVG dataset and ``videoSRC16'' in MCL-JCV dataset, are used to present the progressive decoding feature of our method, as visualized in Fig.~\ref{fig:bosphorus_progressive} and Fig.~\ref{fig:videoSRC16_progressive}. The columns correspond to the frame order, in which the first frame is intra coded while the others are inter coded. The rows are decoded results using the bitstreams at the current and preceding scales. It clearly shows that as more latent variables are used, the reconstruction quality of the corresponding frames gradually improves while consuming more bitrate. A portion of the bitstreams used for decoding of the preceding frames will not impact the decoding of the following frames, and the quality among frames can still be relatively stable. This also explains that our approach provides some resistance to packet loss. As long as the previous frames have a certain level of bitstreams to provide decoding, the subsequent frames can also get at least the same level of decoding results.

\end{document}